\documentclass[11pt,a4paper]{article}

\usepackage{fullpage}
\usepackage{natbib}

\usepackage{amsmath}
\usepackage{authblk}
\usepackage{mathtools}
\usepackage{graphicx}
\usepackage{multirow}
\usepackage{amssymb}
\usepackage{epstopdf}
\usepackage{subcaption}
\usepackage{cancel}
\usepackage{enumitem}
\usepackage{placeins}
\usepackage{algorithm}
\usepackage{algpseudocode}
\usepackage{lineno}
\usepackage[colorlinks, citecolor=blue]{hyperref}

\newcommand{\Dcal}{\mathcal{D}}
\newcommand{\Vcal}{\mathcal{V}}

\newcommand{\Rcal}{\mathcal{R}}

\newcommand{\Scal}{\mathcal{S}}
\newcommand{\Ccal}{\mathcal{C}}

\newcommand{\Acal}{\mathcal{A}}

\newcommand{\Ecal}{\mathcal{E}}
\newcommand{\Mcal}{\mathcal{M}}

\title{Real-world Meeting Points for Shared Demand-Responsive Transportation Systems}

\author[1]{Paul Czioska\thanks{paul.czioska@ikg.uni-hannover.de}}
\author[2]{Ronny Kutadinata\thanks{ronny.kutadinata@unimelb.edu.au}}
\author[3]{Aleksandar Trifunovi{\'c}\thanks{a.trifunovic@tu-braunschweig.de}}
\author[2]{Stephan Winter\thanks{winter@unimelb.edu.au}}
\author[1]{Monika Sester\thanks{monika.sester@uni-hannover.de}}
\author[3]{Bernhard Friedrich\thanks{friedrich@tu-braunschweig.de}}

\affil[1]{Institute of Cartography and Geoinformatics, Leibniz Universit\"at Hannover, Germany}
\affil[2]{Department of Infrastructure Engineering, The University of Melbourne, Australia}
\affil[3]{Institute of Transportation and Urban Engineering, Technische Universit\"at Braunschweig, Germany}

\begin{document}

\maketitle

\begin{abstract}
While conventional shared demand-responsive transportation (SDRT) systems mostly operate on a door-to-door policy, the usage of meeting points for the pick-up and drop-off of user groups can offer several advantages, like fewer stops and less total travelled mileage. Moreover, it offers the possibility to select only feasible and well-defined locations where a safe (de-)boarding is possible. This paper presents a three-step workflow for solving the SDRT problem with meeting points (SDRT-MP). Firstly, the customers are clustered into similar groups, then meeting (and divergence) points are determined for each cluster. Finally, a parallel neighbourhood search algorithm is applied to create the vehicle routes. Further, a simulation with realistic pick-up and drop-off locations based on map data is performed in order to demonstrate the impact of using meeting points for SDRT systems in contrast to the door-to-door service. Although the average passenger travel time is higher due to enhanced walking and waiting times, the experiment highlights a reduction of operator resources required to serve all customers.
\end{abstract}


\textit{Preprint submitted to Transportation Science}

\newpage

\FloatBarrier
\section{Introduction}

Demand-responsive transportation (DRT) services, also known as Dial-a-Ride, provide a mobility solution based on door-to-door transportation on request. They can be operated by companies or statutory authorities. Initially intended as a service with restricted usage (such as for disabled or elderly), it attracted more attention in recent years due to emerging mobility solutions and the shortcomings of conventional public transportation systems \citep{nelson2010recent,Navidi2016}. The DRT trend is further boosted by the rapid developments in information and communication technologies which help to automatically process requests and assign vehicles.

While taxi and conventional DRT services accommodate mostly one customer (or a customer group) at a time, shared demand-responsive transportation services (SDRT) aim at utilising idle resources by combining several requests with similar itineraries and time schedules. Although this induces possible detours to accommodate multiple passengers, the service costs can be reduced and the vehicle capacity utilization can be improved. Growing traffic problems such as congestion or pollution in urban areas encourage a shift towards shared mobility as a more sustainable practice.
In the last decade, many companies have launched new shared demand-responsive transportation services, including popular companies like UberPOOL\footnote{\url{http://www.uber.com/de/ride/uberpool}} and Lyft Line\footnote{\url{http://www.lyft.com/line}}, and smaller local startup companies such as Bridj\footnote{\url{http://www.bridj.com}} (Boston, Kansas), Via\footnote{\url{http://ridewithvia.com/}} (New York, Chicago, Washington D.C), CleverShuttle\footnote{\url{http://clevershuttle.org/}} (Berlin, Leipzig, M\"unchen), and Allygator\footnote{\url{http://www.allygatorshuttle.com/}} (Berlin).

Most of the SDRT services focus on door-to-door transportation, offering passengers the ability to choose the desired pick-up and drop-off location by themselves. However, some recently launched SDRT service providers, such as Uber, Bridj and Via, deviate from this paradigm and operate on a meeting-point-based mode. The passengers are expected to walk to a meeting point recommended by the system and, after the ride, are dropped off at another predefined point, close to their desired destinations. Moreover, multiple passengers can be grouped together and picked-up (and dropped-off) at the same meeting (and divergence) point. This provides the benefit of reducing the number of stops and the service time. \cite{stiglic2015benefits} have shown that, in a ride-sharing system, the use of meeting points also has the ability to improve the percentage of matched participants and to reduce the total mileage of vehicles.

In the literature, the use of meeting points has not gained much attention compared to conventional door-to-door DRT systems. Although there are some simulation studies that investigate the impact of meeting points, the actual determination of eligible meeting points in a real city environment is mostly neglected. Often, the Euclidean plane is used for simulation, or all vertices of the street network are considered as potential meeting point location (e.g. \cite{stiglic2015benefits,balardino2016heuristic,aissat2014dynamic}). In reality, however, suitable locations for a safe and convenient pick-up and drop-off are not ubiquitous, since it may not be possible to stop at a junction or in the middle of a street. Moreover, feasible meeting point candidates, such as public parking areas, are usually unequally distributed within the city area and dissimilarly reachable by vehicles and pedestrians. Further, the road network may contain obstacles and one-way streets that require large detours to reach some meeting points. Hence, the impacts of these limitations need to be investigated regarding a SDRT scenario.

Besides, the naming of meeting points is not consistent in the literature. Frequently used denominations include \emph{meeting point}, \emph{pick-up point}, \emph{boarding point}, \emph{stopping point}, \emph{ride-access point} and \emph{rendezvous point}; and correspondingly: \emph{drop-off point}, \emph{deboarding point} and \emph{leaving point}. For the sake of consistency, the denominations \emph{meeting point} (MP) and \emph{divergence point} (DP) will be used in this paper.

This paper presents a workflow towards a realistic inclusion of meeting points for SDRT services in urban areas. The approach uses real-world meeting points, such as from OpenStreetMap\footnote{\url{http://www.openstreetmap.org/}}. The resulting SDRT with meeting points (SDRT-MP) problem is solved with a multi-step approach. Firstly, the incoming requests are clustered into groups of equal size. In the second step, feasible pick-up and drop-off locations for each cluster are obtained while ensuring acceptable walking distances and time windows. If it is potentially beneficial, multiple alternative meeting point locations are proposed, which are considered within the routing optimisation. Finally, a vehicle routing optimisation algorithm is applied to determine efficient routes with meeting points to serve the customers. Experimental results show that the use of meeting points can offer several benefits especially for the service operator compared to a conventional door-to-door service.

The paper is organized as follows. The review of related works in the literature is presented in the next section, followed by a detailed description of the proposed workflow in Section \ref{sec:workflow}. Subsequently, a simulation experiment (Section \ref{sec:experiment}) is presented, followed by the results (Section \ref{sec:results}) and a discussion (Section \ref{sec:discussion}).

\section{Literature Review}
\label{sec:literature}

This section covers related works to the given problem. Firstly, an introduction to general vehicle routing problems is given, followed by a subsection on clustering techniques. Finally, a review on other works that deal with meeting points is presented. 

\subsection{Capacitated pick-up and drop-off problem with time windows}

The fleet management of a DRT service is commonly referred to as the Dial-A-Ride Problem (DARP). The DARP aims to create an efficient route plan in order to satisfy a set of transportation requests. Each request consists of a set of users, intending to travel from an origin to a destination within associated pick-up and drop-off time windows. As long as the capacity of the vehicles is not exceeded, requests with similar itineraries can share the same vehicle. Typically, hard time windows are assumed where customers must be visited within the nominated time windows for the solution to be feasible.  However, with soft time windows, vehicle routes are penalized but still considered feasible if they arrive at the customer locations outside of the time window. A DARP can be placed into two categories, namely a static case, where all requests are known in advance, and a dynamic case, where some or all of the requests appear on short notice. This paper only focuses on the former, which is essentially the capacitated pick-up and delivery (or drop-off) problem with time windows (CPDPTW). It was defined by \cite{Savelsbergh1995} and is proved to be NP-hard. A survey of the CPDPTW is presented by \cite{Parragh2008}. 

There are a number of exact solution methods for the CPDPTW in the literature, such as dynamic programming \citep{Psaraftis1986,Desrosiers1986}; mixed integer programming \citep{Ropke2009,Baldacci2011} and constraint programming \citep{Zhou2009}. Furthermore, there are several approximation methods, which can be categorised as solution construction techniques \citep{Lu2006,Gronalt2003}, solution improvement approaches \citep{Hasle2007,Bent2006,Ropke2006}, and others that do not fit into these categories \citep{Xu2003,Koning2011}. Popular global optimisation techniques, such as genetic algorithm \citep{Pankratz2005,Hosny2009,Nagata2010} and ant colony optimisation \citep{Badaloni2008,Huang2010}, can also be used to solve the CPDPTW. Arguably, the most common solution method used to solve a CPDPTW is the neighbourhood search heuristic \citep{Bent2004a,Ropke2006,Pisinger2010,Demir2012,Ribeiro2012,Azi2014,Li2016}, sometimes coupled with a column generation method to reduce the scale of the resulting optimisation problem \citep{Feillet2010}.

The use of meeting points in SDRT creates a new type of problem, CPDPTW with meeting points (CPDPTW-MP), and introduces two novel challenges. Firstly, the formulation in the literature needs to be modified to include the possibility of selecting a meeting point from a number of alternatives. Secondly, typical to other optimisation problems, CPDPTW suffers from the curse-of-dimensionality, which is worsened by the introduction of meeting point selection in CPDPTW-MP. A common approach to handle the scalability issue is to decompose the problem into assignment problem and single-vehicle routing problem \citep{Cordeau2007}.

\subsection{Demand clustering}

Clustering techniques are a common heuristic to solve and combat the computational complexity of DARP. In general, there are two clustering approaches used for DARP: vehicle based and spatio-temporal based clustering.

In a vehicle based approach, the requests are clustered, and a single vehicle is assigned to serve each group \citep{Bodin1986, Desrosiers1986, Borndorfer1999, Pankratz2005, Hame2015}. This approach is also commonly referred to as the assignment problem. What follows is multiple single-vehicle routing problems, which can be solved independent of each other. The popularity of this approach is attributed to the fact that it provides a partial solution to the DARP by dividing it into two simpler subproblems. In addition, the solution is often defined over a graph representation of the actual DARP, which implies that there are some spatial, and in some cases temporal, constraints.

The second approach used spatio-temporal constraints to cluster the requests \citep{Cullen1981, Gidofalvi2008, Ma2013, martinez2014formulating, Morris2016}. In the routing optimization phase, each cluster is treated as a single trip/request. Using this approach it is possible to serve two clusters using a single vehicle at the same time. Hence, it allows for a more flexible routing optimisation phase. The clustering technique used in this paper falls into this category. It is based on nearest neighbour calculations akin to those in pattern recognition and data analysis.

\subsection{Meeting points in ride-sharing}
\label{subsec:MeetingPointLiterature}
In contrast to conventional door-to-door services, the use of meeting points can offer some advantages at the cost of additional customer walking. A common assumption is that people will walk on average 400 m without objection, with an acceptable maximum walking distance of 800 m \citep{untermann1984accommodating, hess2012walking, millward2013active}. Although these estimations vary, it can be argued that a pick-up and drop-off at the doorstep is not essential for the general population, except for the disabled or elderly. The longer walking and waiting time can, on the other hand, be sacrificed for a lower transportation price.

There are several reasons why meeting points can be advantageous for a standard DRT service:
\begin{itemize}
	\item The reduction of the number of stops naturally also reduces the overall service time of the vehicles.
	\item An appropriate meeting point can be chosen, providing a safe boarding and a convenient parking.
	\item Easier identification for both the vehicle and the passenger since more precise, non-ambiguous location descriptions can be provided.
	\item The real origin and destination locations of the customers are not necessarily disclosed or can be obfuscated with some techniques due to privacy reasons \citep{aivodji2016meeting,goel2016privacy}.
	\item It encourages walking that increases daily physical activities, the lack of which may lead to severe medical consequences \citep{ewing2003relationship, ewing2014relationship,GilesCorti2016}.
\end{itemize}

In general, meeting points have mostly been studied in the scope of ride-sharing and carpooling, where it is essential to determine a meeting location for a driver/passenger match. For scientific experiments, they are sometimes placed on the Euclidean plane \citep{stiglic2015benefits,hall2008evaluation} or the result of an optimization, e.g. as cluster centroid locations \citep{martinez2014formulating}. More detailed settings use the street network for the MP determination. In a ride-sharing scenario, \cite{balardino2016heuristic} assign passengers and drivers to meeting points based on a real street network, enforcing a maximum detour length of the driver and a maximum vehicle capacity. Once drivers and passengers are paired, \cite{yan2011efficient} present an optimal and a greedy solution method if only one meeting point is needed. Alternatively, \cite{aissat2014dynamic, aissat2015meeting} propose several exact and heuristic methods based on extensive shortest path computations to find meeting and divergence points.

In the literature, there have been mixed results on the impact of meeting points for SDRT. \cite{hall2008evaluation} simulated a DRT as a part of an integrated public transport system, where the meeting points are uniformly placed based on a grid. They discovered that the MP-based operation does not seem to give any major differences in the simulation results compared to a door-to-door service concerning the efficiency. However, they also state that if the operator can select the MPs of the customers, the MP solution may be more beneficial. \cite{stiglic2015benefits} found that, in a ride-sharing scenario based on real-world demand data, the introduction of meeting points can improve several metrics, including the percentage of matched participants and mileage savings. This contradiction demonstrates the need for further investigation on the impacts of MP in SDRT services.

\FloatBarrier
\section{Solution workflow}
\label{sec:workflow}


This section describes the proposed workflow used to solve the \emph{SDRT-MP}. It basically consists of three discrete steps:
\begin{enumerate}
	\item \textbf{Clustering} \label{step:cluster}
	\item \textbf{MP candidates selection} \label{step:MP}
	\item \textbf{Routing Optimisation with final MP selection} \label{step:opt}
\end{enumerate}

\begin{figure}[ht] 
  \begin{subfigure}[b]{0.5\linewidth}
    \centering
    \includegraphics[width=0.95\linewidth]{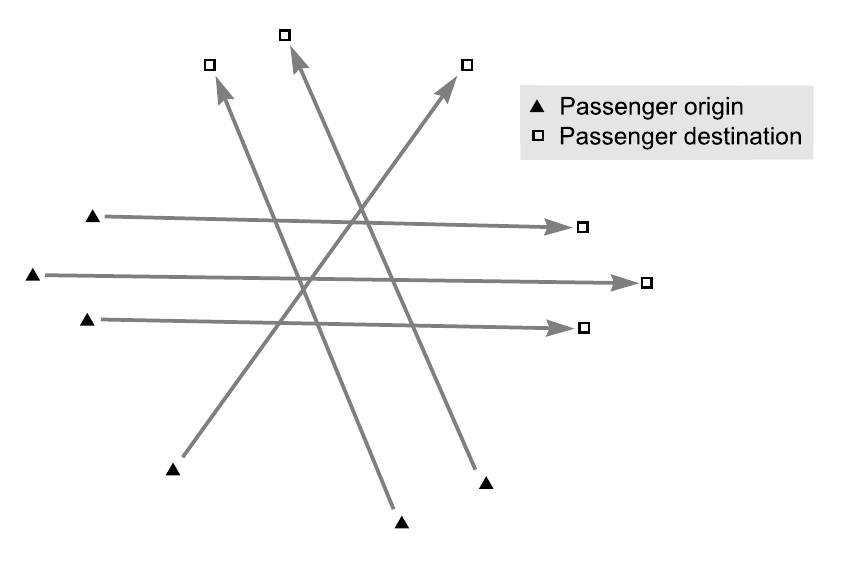} 
    \caption{Initial state with six customers} 
    \label{fig:Workflow_Schema_1} 
    \vspace{1ex}
  \end{subfigure}
  \begin{subfigure}[b]{0.5\linewidth}
    \centering
    \includegraphics[width=0.95\linewidth]{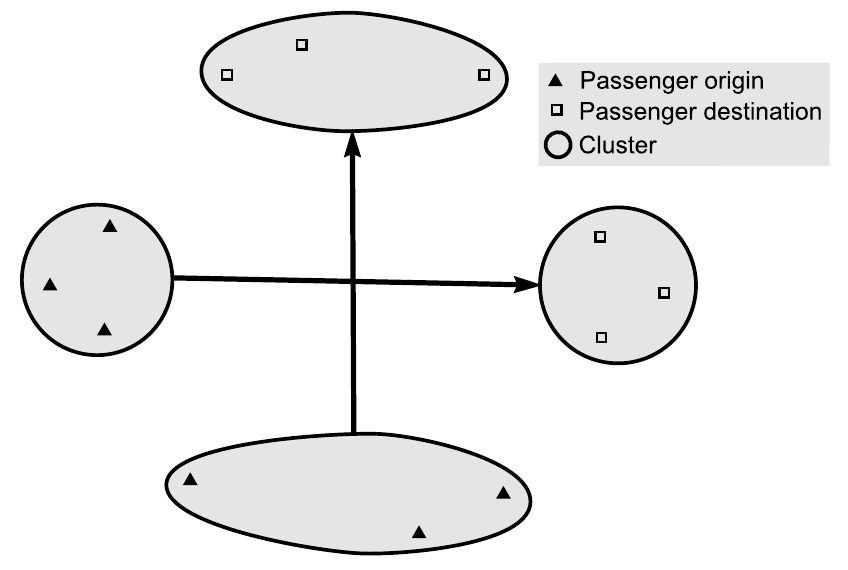} 
    \caption{State after clustering (step \ref{step:cluster})} 
    \label{fig:Workflow_Schema_2} 
    \vspace{1ex}
  \end{subfigure} 
  \begin{subfigure}[b]{0.5\linewidth}
    \centering
    \includegraphics[width=0.95\linewidth]{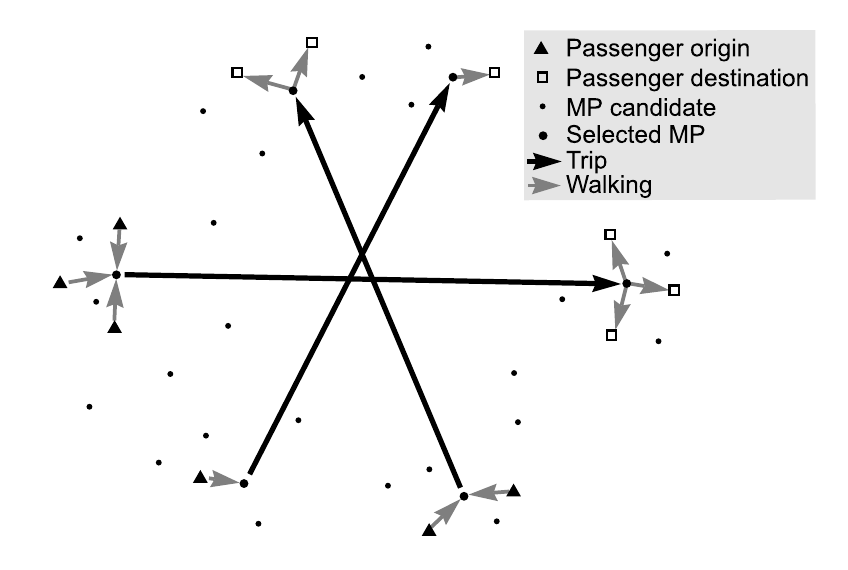} 
    \caption{State after MP candidates selection (step \ref{step:MP})} 
    \label{fig:Workflow_Schema_3} 
    \vspace{1ex}
  \end{subfigure}
  \begin{subfigure}[b]{0.5\linewidth}
    \centering
    \includegraphics[width=0.95\linewidth]{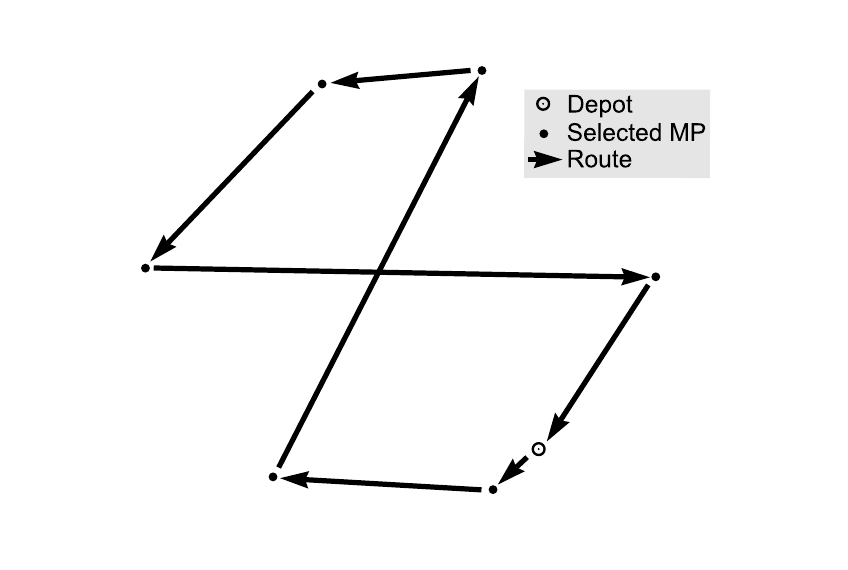} 
    \caption{State after vehicle optimization (step \ref{step:opt})} 
    \label{fig:Workflow_Schema_4} 
    \vspace{1ex}
  \end{subfigure}
  \caption{Basic workflow with clustering, MP selection and route optimization}
  \label{fig:Workflow_Schema} 
\end{figure}

Figure \ref{fig:Workflow_Schema} visualizes the processing chain. In a nutshell, the demand (Figure \ref{fig:Workflow_Schema_1}) is initially clustered into groups with similar itineraries and time schedules (Figure \ref{fig:Workflow_Schema_2}). Secondly, each group is separately split into \emph{trips}, with each trip having a common meeting and divergence point and potential alternative meeting points (Figure \ref{fig:Workflow_Schema_3}). Finally, the vehicle routing problem is solved to construct routes (Figure \ref{fig:Workflow_Schema_4}). All steps are explained in detail in the course of this section.

In this paper, a neighbourhood search approach is applied to solve the vehicle routing optimization (Step \ref{step:opt}). Since this step is very time consuming when being performed on large input data, a modified five-step version is used to enable a parallel processing. For this, the trips resulting from Step \ref{step:MP} are again clustered (\emph{Re-Clustering step}) to partition the trip data into equally large instances. For the Re-Clustering, the same technique as in step \ref{step:cluster} is used. The clustered trips are then processed by the route optimization solver in parallel. Since the partitioned instances are much smaller, the computation time is significantly reduced. Finally, a \emph{Concatenation} step is necessary to merge the different vehicle routes again, which is explained in more detail in Section \ref{subsec:RouteConcatenation}.

In the following, the data prerequisites and the details of each step of the parallel version of the workflow are described.

\FloatBarrier
\subsection{Data prerequisites}
\label{subsec:Prerequisites}

The proposed workflow needs the following input and parameters:
\begin{itemize}
	\item \textbf{Street Network.} A network graph $G = (V, E)$ consisting of vertices $V$ and directed edges $E$. Each edge $e \in E$ has a distance $d(e)$, and a corresponding vehicle driving time $t^{driv}(e)$ and / or a corresponding passenger walking time $t^{walk}(e)$, depending on the type of edge (e.g. footpaths have no defined vehicle driving times).
	\item \textbf{Meeting Point Candidates.} A set of predefined meeting point candidates $\mu \in M \subset V$ and a set of potential divergence point candidates $\delta \in D$. In this paper, these two sets are considered as equal, so that $M = D$. A method to identify the meeting point candidates is presented in Section \ref{subsec:MPcandidate}.
	\item \textbf{Demand.} A set of passenger requests $P$, where each passenger $\rho \in P$ defines an origin vertex $v^{+}$, a desired destination vertex $v^{-}$, and a desired departure time $t^{+}$.
\end{itemize}

\begin{table}[htb]
	\centering
	\caption{Workflow parameters.} \label{tab:parameters}
	\begin{tabular}{|c|c|l|l|}
	    \hline
	    \textbf{Notation}       & \textbf{Unit}     & \textbf{Description}                                      & \textbf{Value used}\\
	                            &                   &                                                           & \textbf{for simulation}\\
	    \hline\hline
	    $t^\text{wait}_{*}$     & s                 & Maximum passenger waiting time                            & 1200 s\\
	    $d^\text{walk}_{*}$     & m                 & Maximum passenger walking distance                        & 800 m\\
        $t^\text{detr}_{*}$     & s                 & Maximum allowed vehicle detour time                       & 1200 s\\
        $r^\text{detr}_{*}$     & \%                & Maximum allowed vehicle detour time percentage            & 25 \%\\
        $t^\text{serv}_{*}$     & s                 & Vehicle service time (for boarding/alighting procedure)   & 120 s\\
        $\sigma$                & \%                & Shortcut ratio threshold (see section \ref{subsec:MPD})   & 50 \%\\
        \hline
        $k_{1}$                 & -                 & Maximum cluster size for the initial clustering           & 11 passengers\\
        $k_{2}$                 & -                 & Maximum cluster size for the re-clustering                & 10 trips\\
        \hline
        $q_{*}$                 & -                 & Maximum vehicle capacity                                  & 9 passengers\\
        $c^\text{dist}_{*}$     & $\text{km}^{-1}$  & Vehicle distance dependent cost                           & 1 / km\\
        $c^\text{vehi}_{*}$     & -                 & Vehicle capital cost                                      & 2000\\
        $c^\text{wait}_{*}$     & $\text{s}^{-1}$   & Passenger wait time cost                                  & 0.5 / second\\
        $c^\text{late}_{*}$     & $\text{s}^{-1}$   & Passenger late time cost                                  & 5 / second\\
        $\alpha$                & -                 & Passenger wait time cost growth                           & 0.5\\
        $\beta$                 & -                 & Passenger late time cost growth                           & 2\\
        \hline
	\end{tabular}
\end{table}
Depending on the maximum passenger walking distance $d^\text{walk}_{*}$, each passenger $\rho \in P$ is able to walk to/from a set of reachable meeting points $M_{\rho}$ and divergence points $D_{\rho}$, respectively. Furthermore, the maximum allowable detour time of a passenger is capped by $t^\text{detr}_{*}$, which is useful to avoid unreasonably long detour on long trips. The resulting maximum detour time is used in the calculation of the latest acceptable arrival time at the destination $t^{-}_{\rho}$ as follows:
\begin{equation}
	\label{eq:LatestArrivalTime}
	t^{-}_{\rho} = t^{+}_{\rho} + t^\text{wait}_{*} + t^\text{serv}_{*} + \min\left(t^{driv}\left(v^{+}_{\rho} \rightarrow v^{-}_{\rho}\right) + t^\text{detr}_{*}, t^{driv}\left(v^{+}_{\rho} \rightarrow v^{-}_{\rho}\right) \cdot r^\text{detr}_{*}\right) \text{.}
\end{equation}

\FloatBarrier
\subsection{Clustering}
\label{subsec:Clustering}

The clustering method in this paper is used to cluster similar transportation requests into groups that are limited in size. It utilises a similar idea to those in the field of data analysis, where abstract points in space are classified using various distance measures in multiple dimension. Although typically maximum spatial and temporal constraints are imposed on clusters, the proposed workflow in this paper applies the constraint checking during the MP candidates selection step (section \ref{subsec:MPD}) to use actual distances based on the street network instead of Euclidean distances.

In a similar fashion to many classification algorithms \citep[chapter~14]{hastie2009elements}, we consider a vector of features $\mathbf{x}=[x_1,x_2,\dots,x_N]^T$, which in this case are the properties of a transport request:
\begin{equation}
\mathbf{x}=[\text{origin coordinates, destination coordinates, desired departure time}]^{T}\text{.}
\end{equation}
The cost function is defined as the Euclidean distance $d(\mathbf{x}_{i}, \mathbf{x}_{k}) = \sqrt{ \sum_{j=1}^{n} (x_{i,j}-x_{k,j})^{2}}$, where $x_{i,j}$ is the $j^\text{th}$ element of vector $\mathbf{x}_i$. To generalize this, consider a cluster $C$ that is a set of transport requests, then the distance of a new transport request $\mathbf{x}$ from cluster $C$ is determined by

\begin{equation}
  d(C, \mathbf{x}):= 
  \begin{cases}
    \sum_{\mathbf{c}\in C} d(\mathbf{c}, \mathbf{x}) &\text{if }  C \neq \emptyset\\
    0 &\text{if } C = \emptyset
  \end{cases}
\end{equation}\label{eq:generalized_distance}

In this paper, the maximum number of requests in the cluster is fixed, so that the created groups are equally sized. The clustering works iterative, so that it processes one request after another. To gain better understanding of this consider a matrix $X$ such that the rows are made of transport requests $\mathbf{x}_i$ where $i$ is the number of rows i.e. requests. In the beginning of the clustering procedure we have $C=\emptyset$. The cost for the first request $x_0$ in $X$ is determined using Equation (\ref{eq:generalized_distance}) which is equal to zero since $C$ is empty. This is the smallest possible cost for the clustering, hence $x_0$ is added to $C$ and $x_0$ is removed from $X$. In the next step, the cost from the cluster $C$ to all other vectors is calculated using Equation (\ref{eq:generalized_distance}) and the vector that has the smallest cost is added to $C$. This is repeated until the maximum allowed size of groups is reached in $C$, after which the whole process is repeated again until all request are processed.

A compact description of the clustering procedure is presented in Appendix~\ref{sec:algorithms} (Algorithm~\ref{alg:clustering}).

\FloatBarrier
\subsection{Meeting Point Candidates Selection}
\label{subsec:MPD}

In this step, feasible meeting and divergence points are assigned to the previously determined clusters. Since the previous clustering step does not use the street network but Euclidean distance, each group may have to be subdivided to satisfy the walking and time window constraints of the customers under real-world conditions. The output of this step is a set of \emph{trips}, each being a tuple of MPs and DPs to be considered in the vehicle routing phase.

\subsubsection{Single MP / DP Selection.}
Firstly, each cluster is divided into trips, each having a single MP and DP. A trip $\tau$ is defined as a tuple consisting of one or more passengers with one or more meeting and divergence point candidates and corresponding time windows, as formula \ref{eq:Trips} shows. $t^{E}$ means here earliest possible arrival, whereas $t^{L}$ indicates the latest possible arrival time at a meeting point.
\begin{multline}
\tau_{i} = \Big( \{\rho_{i,1}, \rho_{i,2}, \cdots\}, 
    \{\left(\mu_{i,1}, t^{E}\left(\mu_{i,1}\right), t^{L}\left(\mu_{i,1}\right)\right), \left(\mu_{i,2}, t^{E}\left(\mu_{i,2}\right), t^{L}\left(\mu_{i,2}\right)\right), \cdots\}, \\
    \{\left(\delta_{i,1}, t^{E}\left(\delta_{i,1}\right), t^{L}\left(\delta_{i,1}\right)\right), \left(\delta_{i,2}, t^{E}\left(\delta_{i,2}\right), t^{L}\left(\delta_{i,2}\right)\right), \cdots\} \Big)
\label{eq:Trips}
\end{multline}

A trip is considered as feasible if the time windows and walking distances are within the thres\-holds for all customers of the trip. The earliest and latest pick-up times for a MP $\mu$ are determined by 
\begin{align}
	t^{E}(\mu) &= t^{+} + t^{walk}(v^{+} \rightarrow \mu) \text{,} \label{eq:TimeWindows_MPE} \\
	t^{L}(\mu) &= t^{+} + t^{walk}(v^{+} \rightarrow \mu) + t^{*}_\text{wait} \label{eq:TimeWindows_MPL}
\end{align}
whereas the earliest and latest drop-off time for a DP $\delta$ are calculated by
\begin{align}
	t^{E}(\delta) &= t^{+} + t^{driv}(v^{+} \rightarrow v^{-}) - t^{walk}(v^{-} \rightarrow \delta) - t^\text{serv}_{*} \text{,} \label{eq:TimeWindows_DPE}\\
	t^{L}(\delta) &= t^{+} + t^\text{wait}_{*} + \min\left(r^\text{detr}_{*} \cdot t^{driv}\left(v^{+} \rightarrow v^{-}\right), t^{driv}\left(v^{+} \rightarrow v^{-}\right) + t^\text{detr}_{*}\right) - t^{walk}\left(v^{-} \rightarrow \delta\right) - t^\text{serv}_{*} \label{eq:TimeWindows_DPL}
\end{align}
respectively.

The stated problem can be interpreted as \emph{set covering} problem, where the task is to find the smallest number of subgroups that satisfy the walking and time constraints of the passengers out of a given cluster. The set covering problem was shown to be NP complete by \cite{karp1972reducibility}; nevertheless, due to the limited size of the clusters resulting from step \ref{step:cluster}, we propose an exact recursive algorithm to solve the given set covering problem.

Initially, the algorithm attempts to put all customers within a cluster into a trip with a common meeting and divergence point. If this is spatially and/or temporally infeasible, the group is split into all possible subgroup combinations, and their feasibility is checked likewise. This is done recursively for each subgroup until a feasible solution is found how the group can be split up. The main objective is to split the initial cluster into as few separated trips as possible. If there are multiple different combinations with the same amount of necessary trips, the combination with the least sum of squared walking distances of all customers is chosen. The squaring is applied to penalize longer walking distances more than short distances to equally distribute the walking distances among the passengers.

As stated above, the method is NP-complete. However, by storing the intermediate results during the recursive process the computation time can be considerably reduced (\emph{dynamic programming} approach). Thus it is possible to process reasonably sized clusters within an acceptable time (see Figure \ref{fig:groupsize_time} for an experimental determination of a limitation). The proposed MP candidates selection method is outlined in Appendix \ref{sec:algorithms} (Algorithm \ref{alg:meetingPoints}).

\subsubsection{Alternative MP / DP Selection.}
\begin{figure}[!ht]
	\centering
	\includegraphics[width=\textwidth]{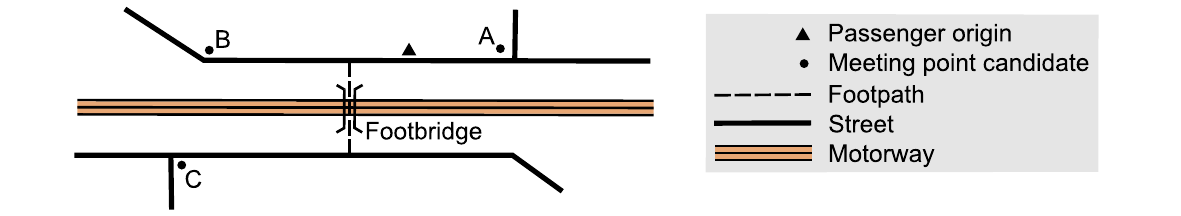}
	\caption{Schematic drawing of a situation with useful alternative meeting point search}\label{fig:AMP}
\end{figure}

The previously described algorithm returns exactly one MP and DP (including time windows) for each trip. However, there can be other meeting points worth considering from the operators perspective. For instance, consider the scenario shown in Figure \ref{fig:AMP}. In this case, the closest meeting point from the passenger origin (Point A) is chosen, which is located north of the motorway. Two other meeting point candidates, namely B and C, would also be feasible for the trip, but they have not been chosen because of longer walking distances. For the operator, however, it could be advantageous in the routing phase to also consider C, since it offers the possibility to approach the passenger from the south without having to take a large detour around the motorway. From the passenger's perspective, it is only a minor extension of the walking path via the footbridge. To combat this, a second algorithm is proposed to identify alternative meeting points.

Using a \emph{shortcut ratio threshold} $\sigma$, an alternative meeting point candidate $\mu_{a}$ is considered if
\begin{equation}
	\frac{t^{driv}(\mu \rightarrow \mu_{a})}{t^{walk}(\mu \rightarrow \mu_{a})} \geq \sigma\text{,} \label{eq:ratioAMP}
\end{equation}
where $\mu$ is the initially identified meeting point candidate. In the example shown in Figure \ref{fig:AMP}, Point C is likely to satisfy (\ref{eq:ratioAMP}), whereas Point B is unlikely. A larger $\sigma$ indicates fewer alternatives, which helps to reduce the input size of the vehicle routing problem. For faster computation times, all ratio values between all meeting point pairs can be precomputed and stored. The proposed algorithm is outlined in Appendix \ref{sec:algorithms} (Algorithm \ref{alg:meetingPointAlternatives}). It recursively checks the next best option until no further useful alternative can be found. Also, note that the corresponding time windows of the alternative meeting points have to be checked. 

\FloatBarrier
\subsection{Route Optimization with Final Meeting Points Selection}
\label{subsec:RouteOptimization}

In this step, the trips having one or more meeting point alternatives resulting from the previous unit are combined and concatenated to vehicle routes. In order to speed up the process, a second clustering is initially applied on the trips to form equally sized trip bunches with similar itineraries, which can then be solved in parallel. For this, the already described clustering method (section \ref{subsec:Clustering}) is reused, however with a different threshold ($k_{2}$). Certainly, this makes it necessary to append a postprocessing step to combine the results of the simultaneously derived vehicle routes (section \ref{subsec:RouteConcatenation}).

The proposed route optimisation problem itself differs from those in the literature mainly because of the alternative meeting points for each boarding/alighting procedure that introduces an extra complexity to the problem. The presented approach of the route optimisation problem follows the formulation by \citep{Kutadinata2017}, which is similar to typical vehicle routing problems \citep{Cordeau2007, Ropke2006, Li2014}.

The problem is formulated as an optimisation model on a directed graph (which is a different graph to the previously defined directed street graph $G$). Firstly, the stops to be visited are represented by a set of nodes, which include the vehicle starting points, meeting and divergence points, and a depot. Note that although each stop can have multiple alternative meeting points, it is still represented by one node. The alternative meeting points modelling will be explained later in this subsection. 

Although the vehicles are located at the depot at the beginning, the formulation differentiates between the starting points and the depot to accommodate the vehicle capital cost (which will become obvious later in the formulation). Next, the index set $\Vcal=\{0,\Scal,\Mcal,\Dcal\}$ is assigned to these nodes, where $\Scal$ is the set of $\nu$ starting nodes of vehicles, $\Mcal$ is the set of $n$ meeting point vertices, $\Dcal$ is the set of corresponding $n$ divergence point vertices, Node $0$ is the depot, and a demand is represented by a pair of meeting and divergence points $(i,n+i)$. These nodes are connected with directed edges, where each edge indicates a possible route to be traversed by the vehicles. Let $\Ecal$ be defined as the set of all directed edges in the network. A directed edge $\epsilon_{i,j}\in\Ecal$ connects a pair of nodes, from Node $i$ to Node $j$, that belongs to one of the following subsets: for each $i\in\Scal$, $j \in \{0\} \cup \Mcal$; for each $i\in\Mcal$, $j\in\Mcal\cup\Dcal$; for each $i\in\Dcal$, $j\in\Vcal\setminus \{\Scal\cup i-n \}$.

Each node is associated with several parameters that define the trip demand. Similarly, each directed edge is associated with distance and travel time parameters. To take into account the multiple alternative meeting points, some of these parameters are functions of the chosen meeting points. For each $i\in\Mcal\cup\Dcal$, let $N_i$ be the number of alternative meeting points for node $i$. Thus, define $\Acal = [m_1 \,\,\, m_2 \,\,\, \dots \,\,\, m_{\nu+2n+1}]$, where $m_i\in\{1,2,\dots,N_i\}$ is the chosen meeting point alternative for node $i$. Having established this, all the graph parameters can be defined as follows.
Let $[t^{E}_{i}(m_i),t^{L}_{i}(m_i)]$ be the associated time-window of node $i$ when $m_i$ is chosen. Note that the time windows are treated as soft constraints. Furthermore, let $d_{i,j}(m_i,m_j)$ and $t_{i,j}(m_{i},m_{j})$ be the travel distance and time from node $i$ to node $j$. In addition, each node has an associated service time $s_i$ and a load $q_i$. For $i\in \{0\} \cup \Scal$ (the depot and the starting points), $s_{i}=q_{i}=0$, $N_{i}=1$, and $[t^{E}_i(1),t^{L}_i(1)]=[0,\infty]$. For the other nodes, $s_i = t^\text{serv}_{*}$. Moreover, in order to keep track of the vehicle loads, let $q_i^{k}$ be the passenger load on-board vehicle $k$ when it leaves node $i$.

Finally, the optimisation formulation is ready to be presented and the decision variables can be defined. Let $u_i^k$ denotes the time vehicle $k$ starts servicing node $i$. The starting point of each vehicle $k$ is similarly denoted with the index $k$ and $s_k=0$. Hence, the variable $u_k^k$ (which technically is the start of the service time of vehicle $k$ at node $k$) indicates the first departure time of vehicle $k$ from its starting point. Similarly, $u_0^k$ represents the final arrival time at the depot. The binary variable $x^k_{i,j}$ is defined to decide whether vehicle $k$ traverses from node $i$ to node $j$. Thus, the decision variables are the service time $u_i^k$, the binary variable $x^k_{i,j}$, and the meeting points $\Acal$. The optimisation problem is formulated as follows:
	\begin{align}
		\min_{x,u,\Acal} \quad \sum_{i\in\Mcal\cup\Dcal} &\left( c^\text{wait}_{*} \max\{u_i^k - t^{E}_i(m_i), 0\}^{\alpha} + c^\text{late}_{*} \max\{u_i^k - t^{L}_i(m_i), 0\}^{\beta} \right) \nonumber\\
		&+ \sum_{k\in\Scal} \left( c^\text{vehi}_{*} \,\mathrm{sgn}\left(u^k_0\right) + \sum_{i,j\in \Vcal} c^\text{dist}_{*} x^k_{i,j} d_{i,j}(m_i,m_j)  \right) \label{eq:costFunc}
	\end{align}
	subject to:
	\begin{align}
		&\sum_{k\in\Scal} \left( \sum_{j\in\Vcal} x_{i,j}^k \right) = 1 &&  \forall i\in\Mcal \text{,} \label{eq:const_pickuponce} \\
		&\sum_{j\in\Vcal} x_{i,j}^k=\sum_{j\in\Vcal} x_{n+i,j}^k && \forall i\in\Mcal, \forall k\in \Scal \text{,} \label{eq:const_getsdropoff} \\
		&\sum_{i\in\Vcal} x_{k,i}^k = \sum_{i\in\Vcal} x_{i,0}^k = 1 && \forall k\in\Scal \text{,} \label{eq:const_startfinishdepot} \\
		&\sum_{j\in\Vcal} x_{i,j}^k=\sum_{j\in\Vcal} x_{j,i}^k && \forall i\in \Mcal\cup\Dcal, \forall k\in\Scal \text{,} \label{eq:const_leaveanode} \\
		& u_j^k\geq \max\left\{ u_i^k + s_i + t_{i,j}(m_i,m_j) , t^{E}_{j}(m_j) \right\} x_{i,j}^k && \forall i,j\in\Vcal, \forall k\in\Scal \text{,} \label{eq:const_arrival} \\
		& u_{i+n}^k\geq u_i^k && \forall i\in\Mcal, \forall k\in\Scal \text{,} \label{eq:const_pickupbeforedropoff} \\
		& q_j^{k} = (q_i^{k} + q_j) x_{i,j}^k && \forall i,j\in\Vcal, \forall k\in\Scal \text{,} \label{eq:const_loadcumsum_pass} \\
		& 0\leq q_i^{k} \leq q_{*} && \forall i\in\Vcal, \forall k\in\Scal \text{,} \label{eq:const_passload}
	\end{align}
%
where $\mathrm{sgn}(\cdot)$ is a function that returns the sign of the input scalar. For a further parameter description we also refer to table \ref{tab:parameters}.

The formulation ensures that each customer is picked-up only once and is dropped-off at the destination by imposing (\ref{eq:const_pickuponce}) and (\ref{eq:const_getsdropoff}). Furthermore, a vehicle has to start at its corresponding starting point and ends its route at the depot as enforced by (\ref{eq:const_startfinishdepot}) and (\ref{eq:const_leaveanode}). Constraint (\ref{eq:const_arrival}) provides a lower bound on the arrival time of a vehicle at a node and (\ref{eq:const_pickupbeforedropoff}) ensures that the drop-off occurs after the corresponding pick-up. Note that $u_i^k$ can be zero if vehicle $k$ never visits node $i$. Finally, (\ref{eq:const_loadcumsum_pass})--(\ref{eq:const_passload}) are constraints for the passengers load of each vehicle.

The first summation in the objective function is the ``service level cost'', which takes into account the passengers late time, pick-up wait time, and detour time. The parameters $\alpha$ and $\beta$ are introduced to enable various polynomial forms of penalty terms. For instance, a quadratic term ($\alpha=2$) can be used to penalize longer wait/late times more than short wait/late times.

The first term in the second summation is the capital cost of each vehicle used. In a typical scenario, the vehicles start and end at the same depot. Due to (\ref{eq:const_startfinishdepot}), if a vehicle $k$ never leaves the depot (i.e.\ is never used), the solution of the formulation would still indicate that the vehicle ``travels'' back to the depot ($x^k_{k,0}=1$), however with no costs assigned. Finally, a route length minimisation is taken into account by the last term of the objective function.

With this formulation, the time windows are treated as penalty terms instead of hard constraints, which is different to typical formulations in the literature \citep{Bent2004a,Cordeau2007,Ropke2009,Baldacci2011}. This allows the optimisation algorithm to choose a solution that has late services that are justified by the savings in other aspects. Typically, higher penalty weight parameters are used to avoid unreasonable number of late arrivals. However, if hard time window constraints are desired, simply choose very large penalty values. Finally, a two-layer neighbourhood search algorithm is used to solve the optimisation. The top layer optimises the trip allocation to vehicles, which repeatedly calls the bottom layer that optimises the route of each vehicle including the meeting and divergence point selection. The algorithms are presented in Appendix \ref{sec:algorithms} (Algorithms \ref{alg:route} and \ref{alg:trip}).

\FloatBarrier
\subsection{Route concatenation}
\label{subsec:RouteConcatenation}

The output of the optimisation process in the previous step is a group of routes, each route performed by a vehicle. Since the optimisation of the previous step can be performed in parallel for each cluster of trips, some of the routes can now be concatenated to reduce the total number of routes (and consequently the total number of vehicles used). Thus, this step can be described as a problem of maximising the number of concatenations by using a Linear Programming (LP) approach. To ensure that a concatenated route can still be feasibly served by a vehicle, a constraint is applied to ensure that there is enough time to travel from the last stop of a precedent route to the first stop of the subsequent one.

Let $R$ be the set of all routes produced by the route optimisation process. A route $r_i\in R$ is defined by the tuple $\{t^s_i,b^s_i,t^e_i,b^e_i\}$, where $t^s_i$ is the start of the service time of the first stop of $r_i$ (that is not the vehicle starting point), $b^s_i$ is the location of the first stop of $r_i$, $t^e_i$ is the end of the service time of the last stop of $r_i$ (that is not the depot), and $b^e_i$ is the location of the last stop of $r_i$. Moreover, $t^{driv}(b^e_i \to b^s_j)$ is the vehicle travel time from the last stop of $r_i$ to the first stop of $r_j$. Furthermore, define the decision variable $x_{i,j}$, which is equal to one if $r_j$ is appended to the end of $r_i$, and zero otherwise. The optimisation aims at maximising the number of concatenation as follows:

\begin{equation}
	\max_{x} \quad \sum_{i,j} x_{i,j} \label{eq:costFuncLP}
\end{equation}
subject to:
\begin{equation}
	t^e_i + t^{driv}(b^e_i \to b^s_j) \leq t^s_j \text{,}\quad\quad \forall i,j\text{.}
\end{equation}

\begin{figure}[!ht]
	\centering
	\includegraphics[width=0.7\textwidth]{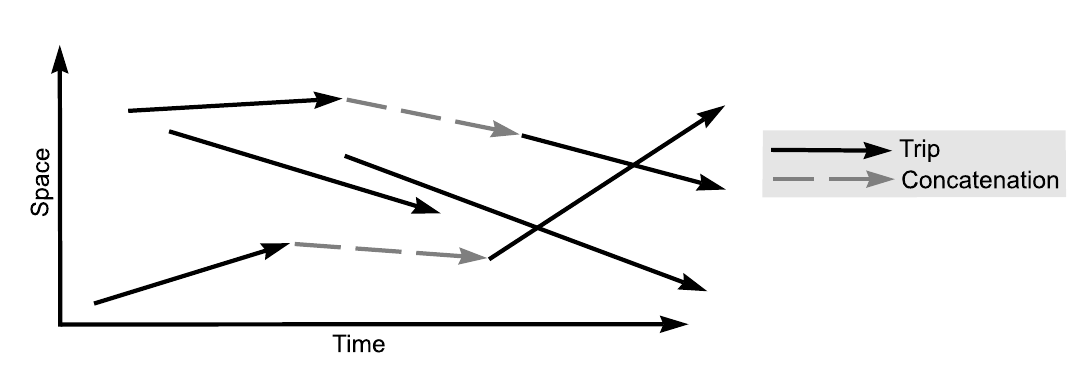}
	\caption{Concatenation of routes to form longer ones}
	\label{fig:Concatenation}
\end{figure}

The proposed concatenation approach is also visualized as a simplified 2D version in Figure \ref{fig:Concatenation}. Note that the final statistics presented in this paper are the output of the LP problem.

\FloatBarrier
\section{Simulation experiment}
\label{sec:experiment}


To evaluate the potential benefits of the use of meeting points as proposed by the workflow, a simulation experiment is carried out, comparing the performance of a meeting-point based service (MP) with a conventional door-to-door service (DS) used as a baseline scenario. In the MP service, all steps of the proposed workflow are executed, whereas the DS service only uses the vehicle routing optimization, applied on the raw demand data. As a result, this simulation focuses on highlighting the benefits of Steps \ref{step:cluster} and \ref{step:MP} rather than the route optimisation itself. Note that for the simulation experiments, the parallel version of the route optimization is used in order to speed up the computation and solve the instances within a reasonable CPU time.

\subsection{Simulation setup}
\label{subsec:SimulationSetup}
\FloatBarrier

The simulation experiment investigates the impact of using meeting points for various demand densities. It is expected that, as the demand density increases, the impact of using meeting points is more significant. 
To this end, a total of 10,000 passengers are randomly generated based on the procedure described in Section \ref{sec:random_demand} and subdivided into four demand instances (1000, 4000, 7000 and 10,000). Each demand instance is subjected to both the MP and DS workflow. To simplify the experiment, only a static problem is considered, i.e. all trips are known in advance. The parameters used in the optimisation and meeting point algorithm are shown in Table \ref{tab:parameters}.

\begin{figure}[!ht]
	\centering
	\includegraphics[width=0.8\textwidth]{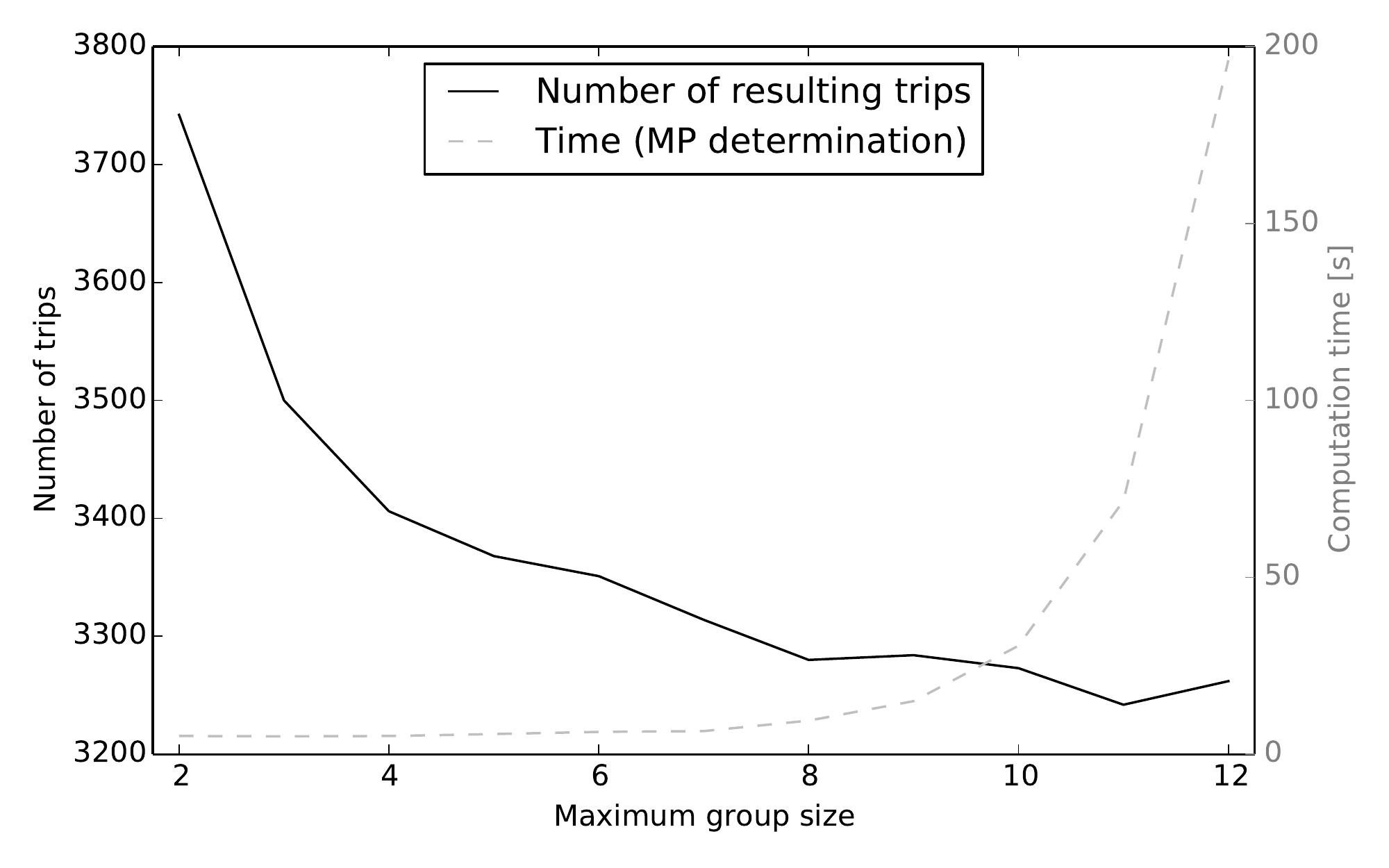}
	\caption{The number of trips and the computation time for different maximum cluster size settings}
	\label{fig:groupsize_time}
\end{figure}

The maximum cluster size for the initial clustering is a tuning parameter that balances the trade-off between the quality of the result and the computation time. Larger clusters lead to better results since the meeting point determination is exact and yields the optimal solution, but the computation time grows exponentially. In order to choose a reasonable maximum cluster size, different values are investigated. Figure \ref{fig:groupsize_time} shows an experiment based on 5000 customers. The maximum cluster size is varied from 2 to 12; the resulting number of trips and the computation times are recorded using parallel processing on 4 cores. Based on this result, a maximum cluster size of 11 has been chosen for the simulation experiment, as it provides good solutions at still reasonable computation times.

\FloatBarrier
\subsection{Street Network}
\label{subsec:StreetNetwork}

For the simulation experiment, the city of Braunschweig, Germany, is used as spatial template. It is a medium-sized city with 250,000 inhabitants and a typical European city structure: The centre is dominated by its historical core with an irregular street network and pedestrian precincts, surrounded by a densely populated area with a more regular street network. In the outskirts, the population density is significantly lower and there are some industrial areas.

\begin{figure}[t]\vspace*{4pt}
    \centering
        \begin{subfigure}[b]{0.45\linewidth}
                \centering
				\includegraphics[width=\linewidth]{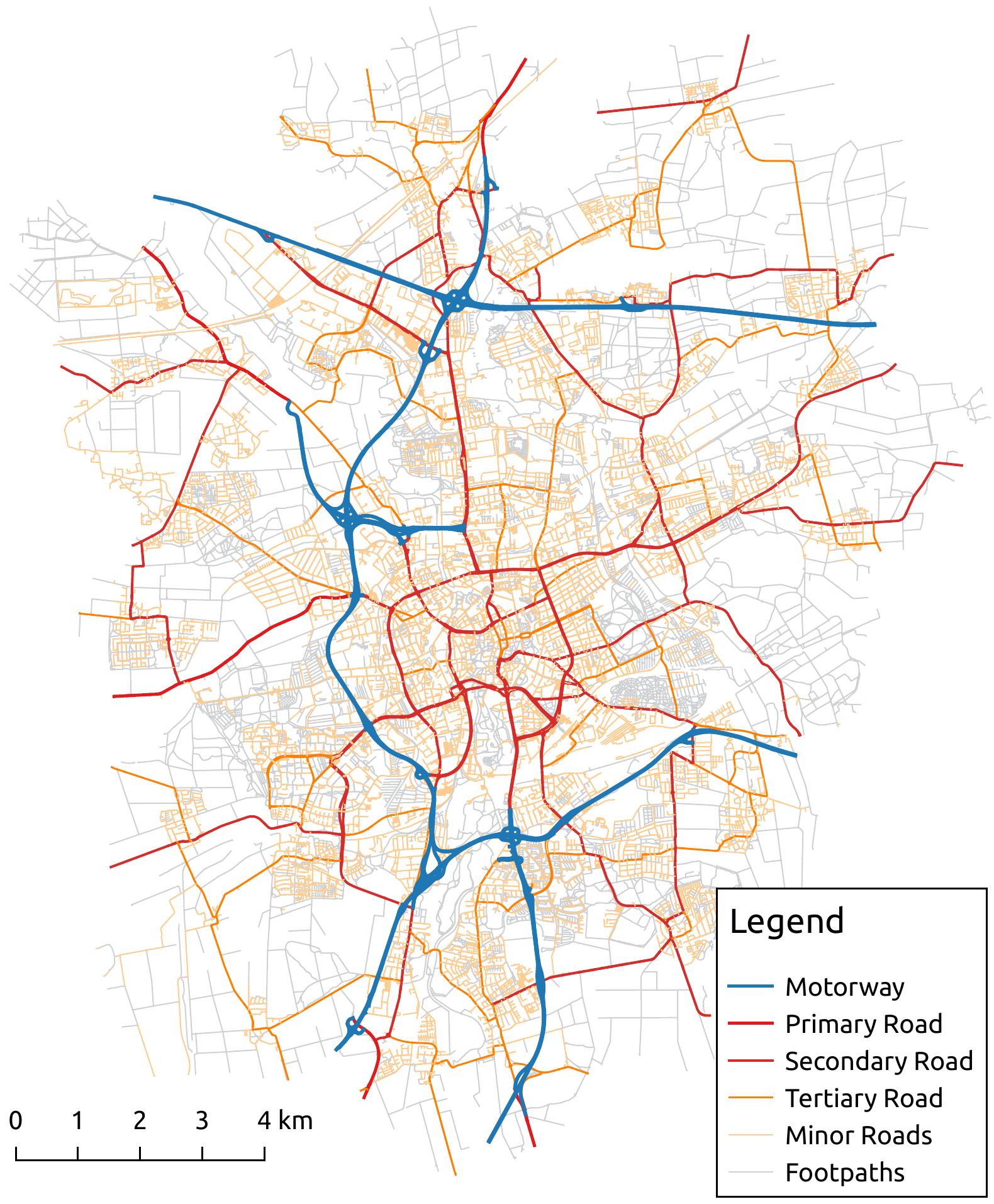}
                \caption{Routing network (Source: OpenStreetMap)}
                \label{fig:BS_Map2}
        \end{subfigure}
        \qquad
        \begin{subfigure}[b]{0.45\linewidth}
                \centering
				\includegraphics[width=\linewidth]{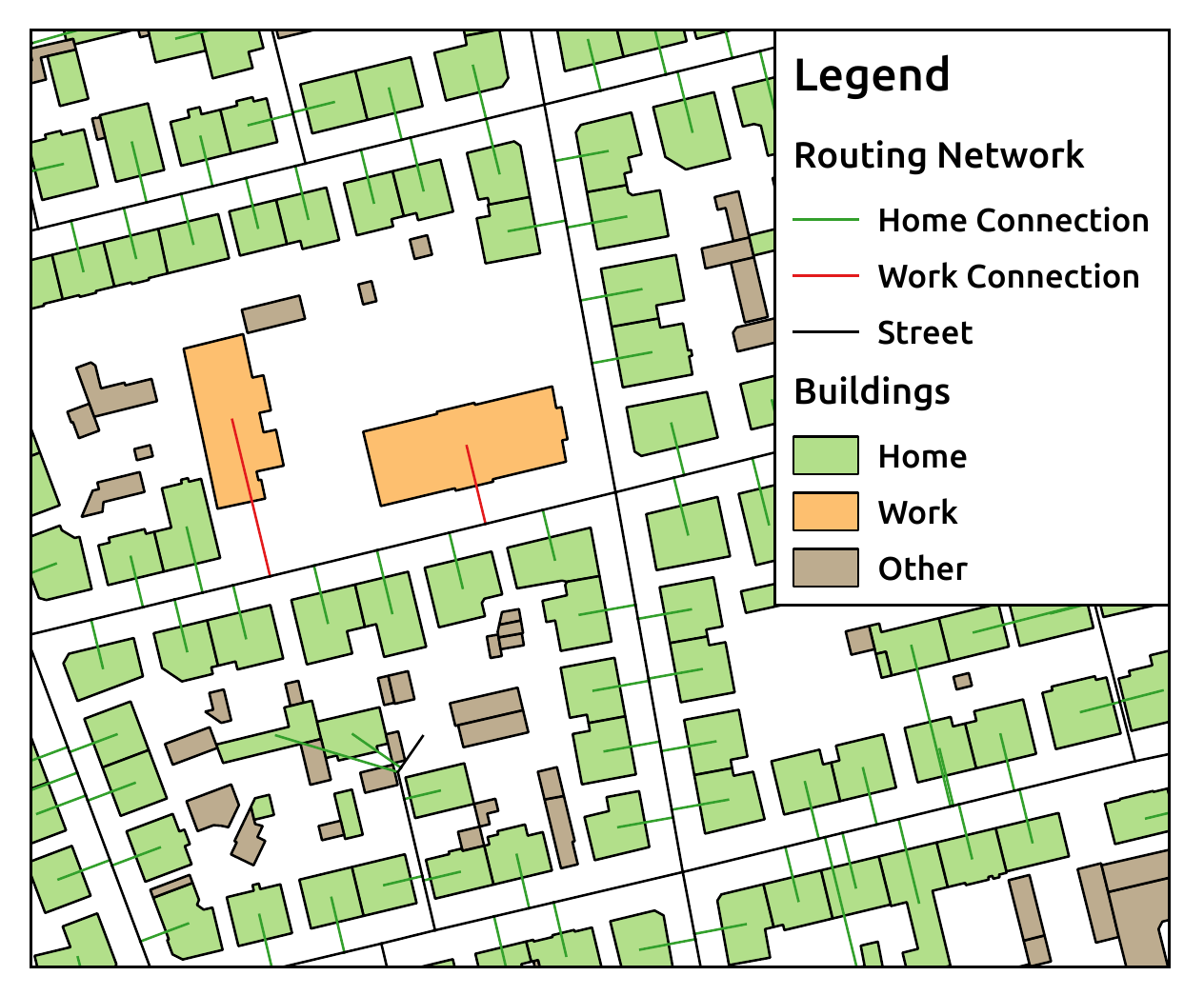}
                \caption{Connected origins and destinations}
                \label{fig:BS_Origins}
        \end{subfigure}
    \caption{Investigation area: City of Braunschweig, Germany}
    \label{fig:BS_Map}
\end{figure}

The street network has been obtained from OpenStreetMap\footnote{\url{http://www.openstreetmap.org/}} and transformed into a routing-enabled graph (Figure \ref{fig:BS_Map2}). As previously mentioned in Section \ref{subsec:Prerequisites}, vehicle driving times $t^{driv}(v_{i} \rightarrow v_{j})$, passenger walking times $t^{walk}(v_{i} \rightarrow v_{j})$ are calculated for every edge $e \in E$. The driving time depends on the maximum allowed speed. The walking time $t^{walk}_{e}$ is determined by assuming a constant walking speed of $4.8$ km/h, using the mean speed for active-transport walking trips revealed by \cite{millward2013active}. The traversal of footpaths, cycle ways and stairways is prohibited for vehicles. Likewise, pedestrians are not allowed to walk on major roads or motorways.

As potential origins, residential buildings from a Level-of-Detail 1 building model of the city of Braunschweig \citep{braunschweigOpenData} with a size of more than 100 $m^{2}$ are added to the network. Likewise, workplace buildings are used for potential destinations. All considered buildings are connected to the street network to model the whole path of the user (Figure \ref{fig:BS_Origins}). In total, the network consists of 88381 nodes and 99497 edges, including 26845 potential home and 2615 potential work locations.

\FloatBarrier
\subsection{Meeting point candidates} 
\label{subsec:MPcandidate}

Feasible MP candidates $\mu \in M$ (and, consequently, also DP candidates) are identified automatically in the map data obtained from OpenStreetMap. In order to ensure safety and convenience aspects like boarding places with reduced traffic, possibilities to park and easily recognizable places, the candidate locations are limited to the following selection:
\begin{itemize}
\item publicly accessible parking places without parking fees,
\item side road intersections (with all adjacent roads having a maximal speed of $\leq$ 30 km/h),
\item turning areas (mostly at the end of a cul-de-sac),
\item petrol stations.
\end{itemize}

All geoobjects are extracted by a GIS workflow from an OpenStreetMap file. Parking areas and petrol station areas are converted to point features using the centroid. Each identified candidate is then connected to the street network $G$ with an edge between the meeting point location and the closest point on the closest edge. If the closest edge is not reachable by vehicles (such as a footpath), a second edge is inserted to the closest drivable edge. The same procedure is applied for edges not accessible by foot. In total, 3475 meeting points have been detected within the investigation area. In relation to the total investigation area size of 193 $km^{2}$, the overall MP density is approximately 18 per $km^{2}$, and in dense urban areas up to 40 MP per $km^{2}$. The observed mean distance to the nearest neighbouring MP is approximately 70 meters.

Compared to the total number of nodes in the graph (excluding house connections) with approximately 30 000 nodes, the number of candidate locations is relatively low compared to other works that use all street nodes as candidate locations, like \cite{balardino2016heuristic}.

\FloatBarrier
\subsection{Random demand}
\label{sec:random_demand}

\begin{figure}[ht] 
  \begin{subfigure}[b]{0.5\linewidth}
    \centering
    \includegraphics[width=0.95\linewidth]{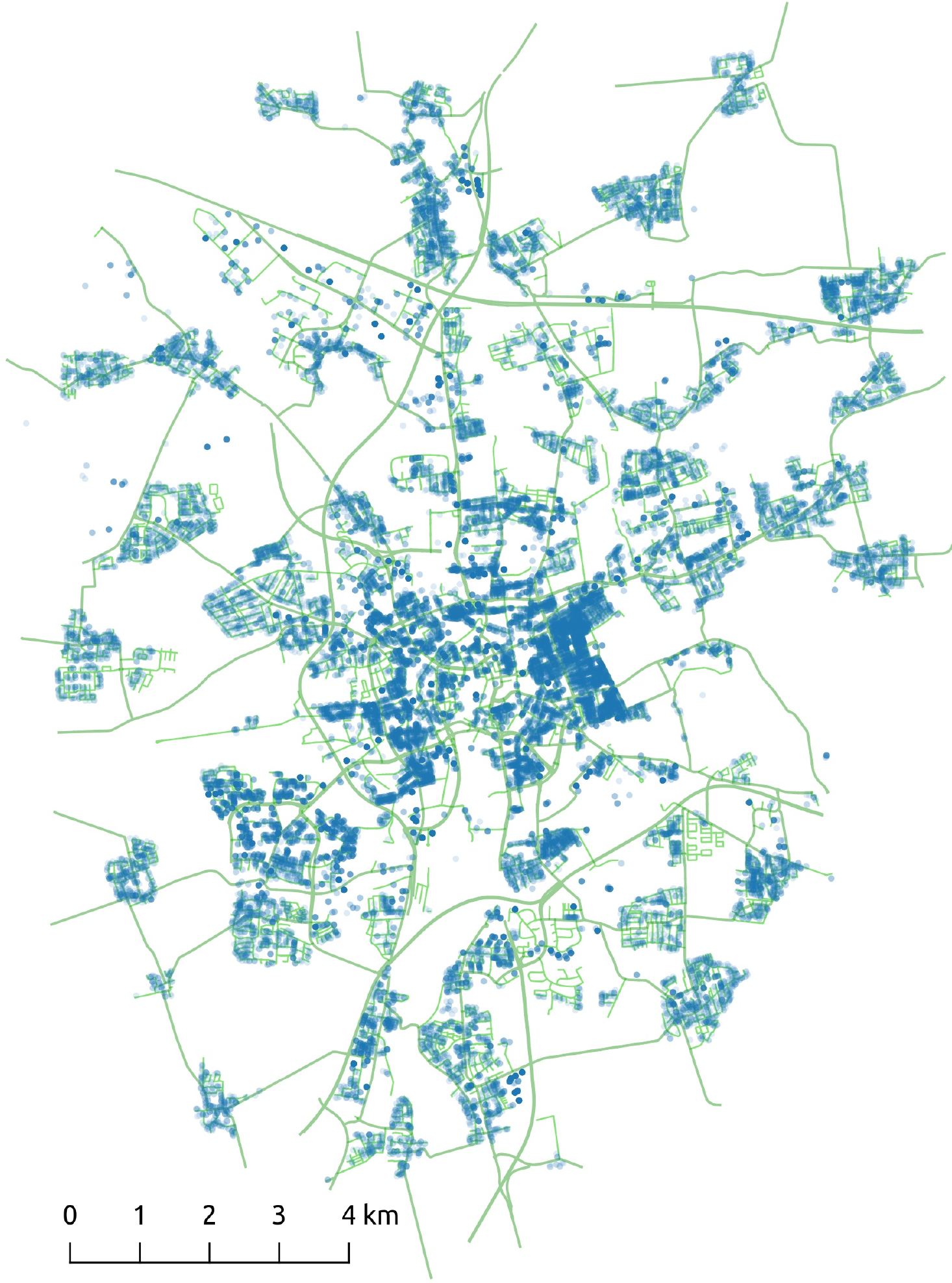} 
    \caption{Customer origin locations} 
    \label{fig:demand_origins} 
    \vspace{1ex}
  \end{subfigure}
  \begin{subfigure}[b]{0.5\linewidth}
    \centering
    \includegraphics[width=0.95\linewidth]{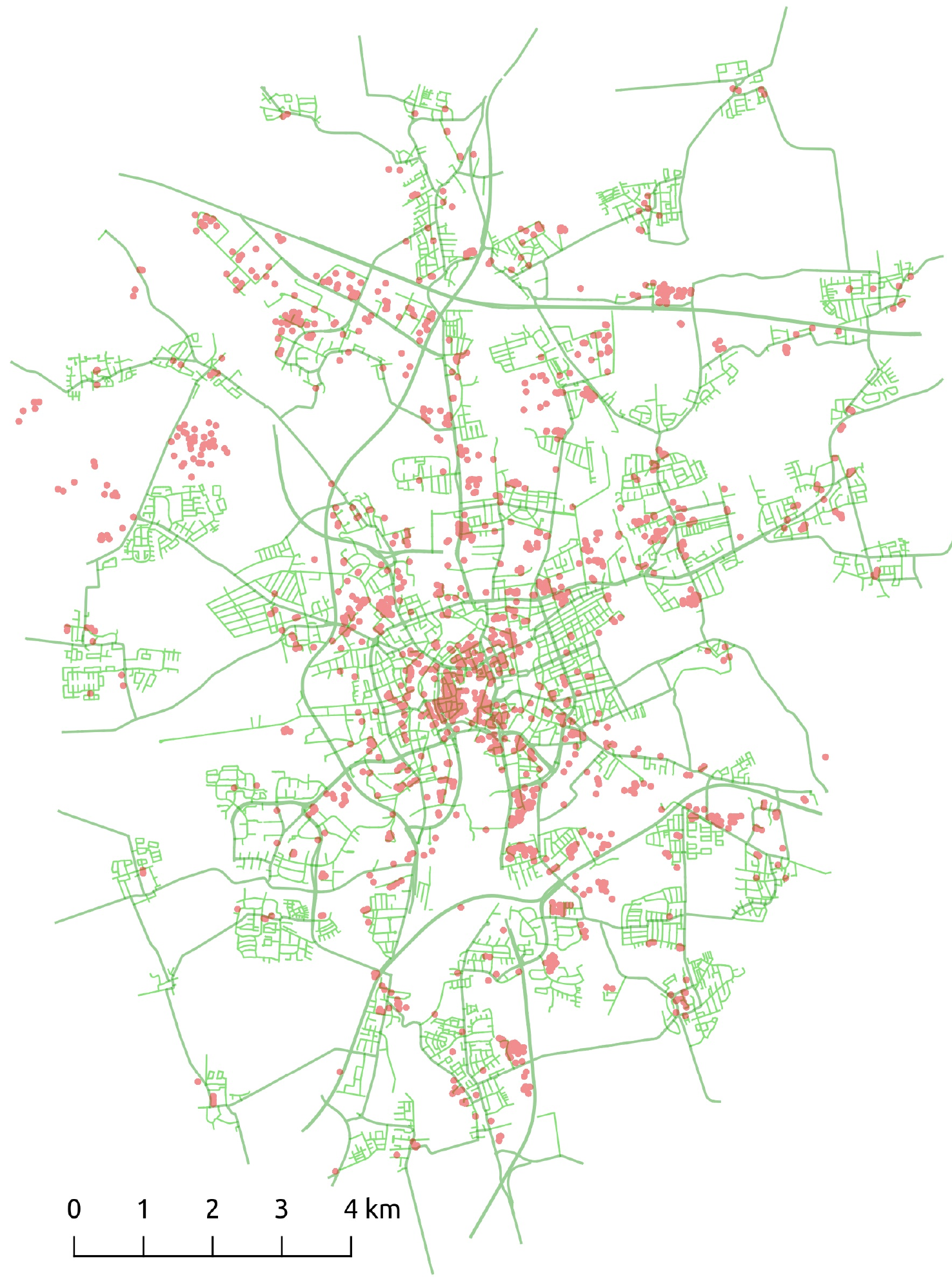} 
    \caption{Customer destination locations} 
    \label{fig:demand_destinations} 
    \vspace{1ex}
  \end{subfigure}
  \caption{Distribution of customer origins and destinations within the investigation area}
  \label{fig:demandlocations} 
\end{figure}

A set of randomly generated passenger requests $P$ is used, defining for each passenger $\rho \in P$ an origin node $v^{+}_{\rho}$, a desired destination node $v^{-}_{\rho}$, and a desired departure time $t^{+}_{\rho}$. For the DS case, the closest node that is reachable by vehicles is stored as the origin, and for the destination likewise. The spatial distribution of the trip requests is based on a weighted sampling of building data. Specifically, the probability of a building to be chosen as origin or destination is proportional to its volume. 
To prevent huge factory buildings to be chosen disproportionately often, the probability of a building being selected is capped equivalent to having a volume of 10000 $m^{3}$. Figure \ref{fig:demandlocations} visualizes the spatial distribution of customers within the investigation area. 

\begin{figure}[ht] 
  \begin{subfigure}[b]{0.5\linewidth}
    \centering
    \includegraphics[width=0.95\linewidth]{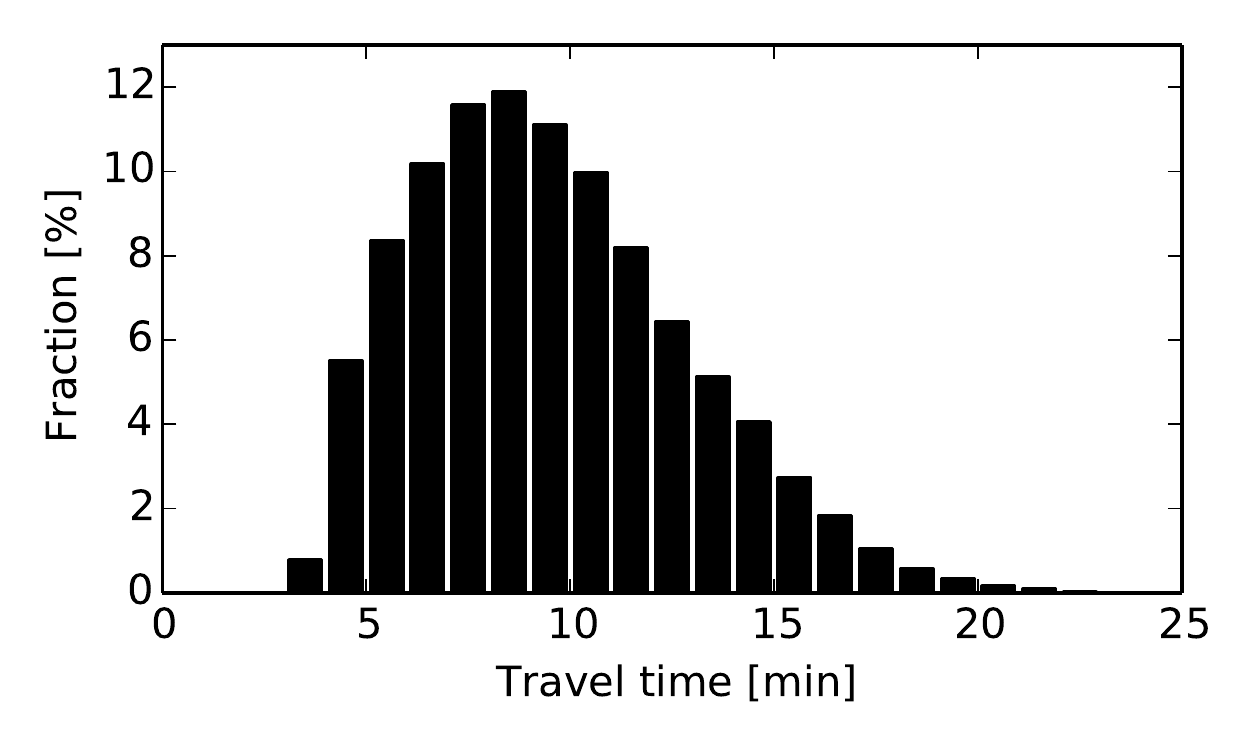} 
    \caption{Distribution of direct vehicle travel times} 
    \label{fig:demand_traveltime} 
    \vspace{1ex}
  \end{subfigure}
  \begin{subfigure}[b]{0.5\linewidth}
    \centering
    \includegraphics[width=0.95\linewidth]{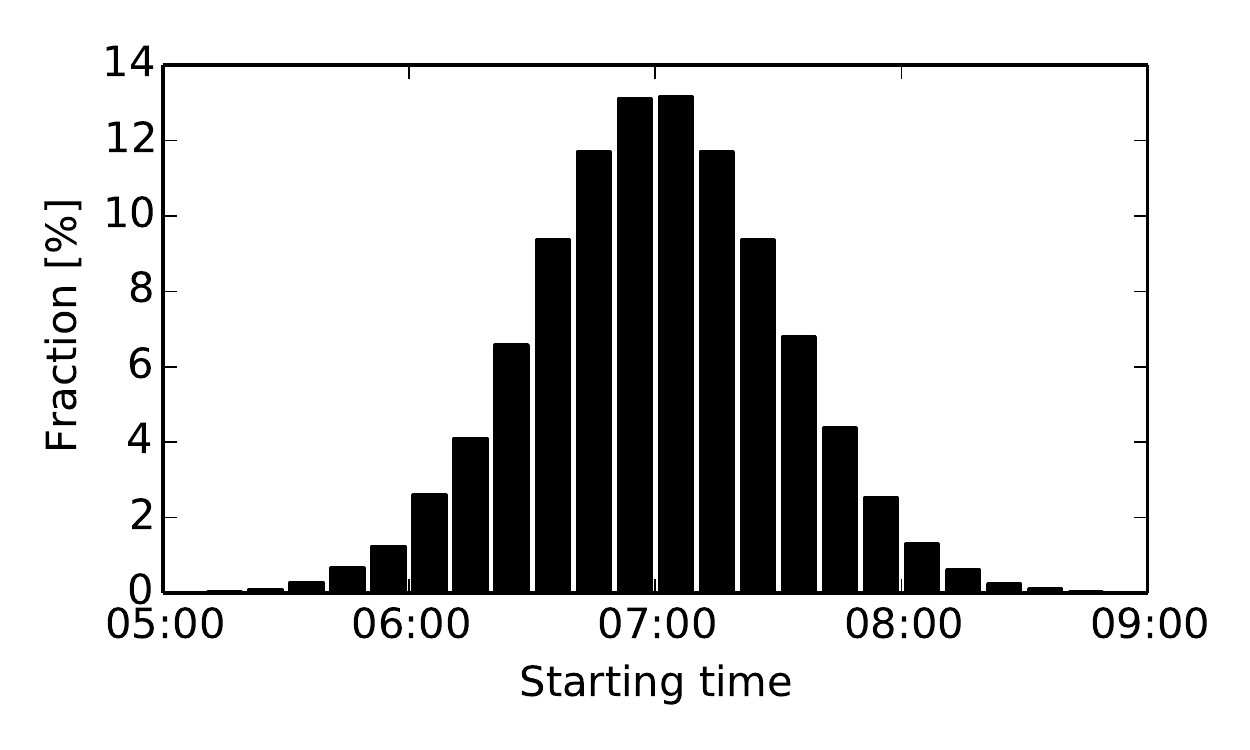} 
    \caption{Distribution of starting times} 
    \label{fig:demand_starttime} 
    \vspace{1ex}
  \end{subfigure}
  \caption{Statistics about randomly generated customer demand}
  \label{fig:demand} 
\end{figure}

Furthermore, trip requests having their destination within 2000 m Euclidean distance of the origin are avoided since the passengers are assumed to walk or cycle. Figure \ref{fig:demand_traveltime} shows the resulting distribution of direct travel times via driving between origins and destinations.

In order to simulate a busy morning commute peak, the temporal distribution of the requests follows a Gaussian distribution centred at 07:00 am with a standard deviation of 30 minutes (Figure \ref{fig:demand_starttime}). Moreover, as there are no actual depots in this problem, the depot location is chosen as the node which has the smallest maximum travel time to the other nodes in the network graph.

\FloatBarrier
\section{Results}
\label{sec:results}

\begin{figure}[!ht]
	\centering
	\includegraphics[width=0.95\textwidth]{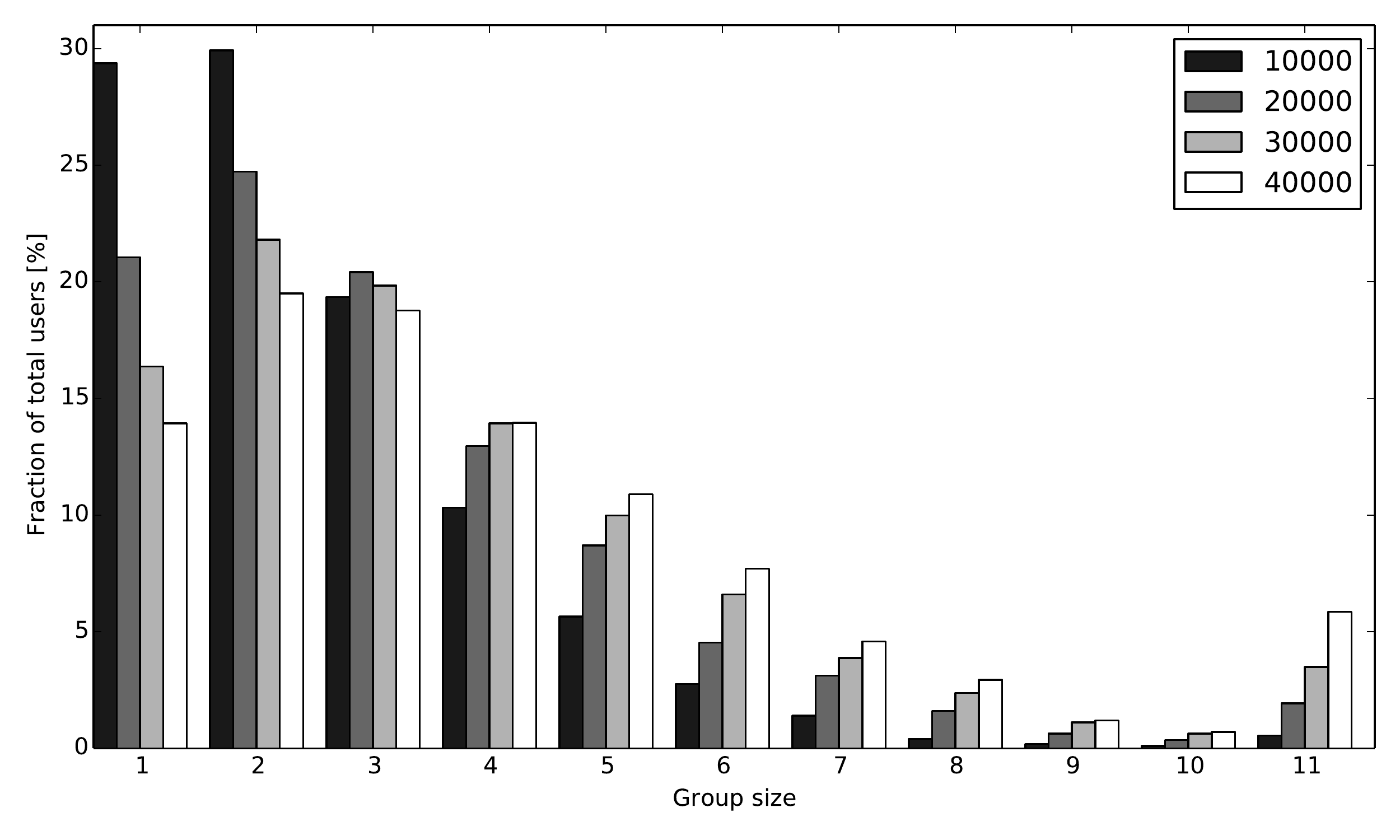}
	\caption{Fraction of users per group size after MP candidates selection for different total demand scenarios}
	\label{fig:groupsize}
\end{figure}

\begin{figure}[ht] 
  \begin{subfigure}[b]{0.5\linewidth}
    \centering
    \includegraphics[width=0.95\linewidth]{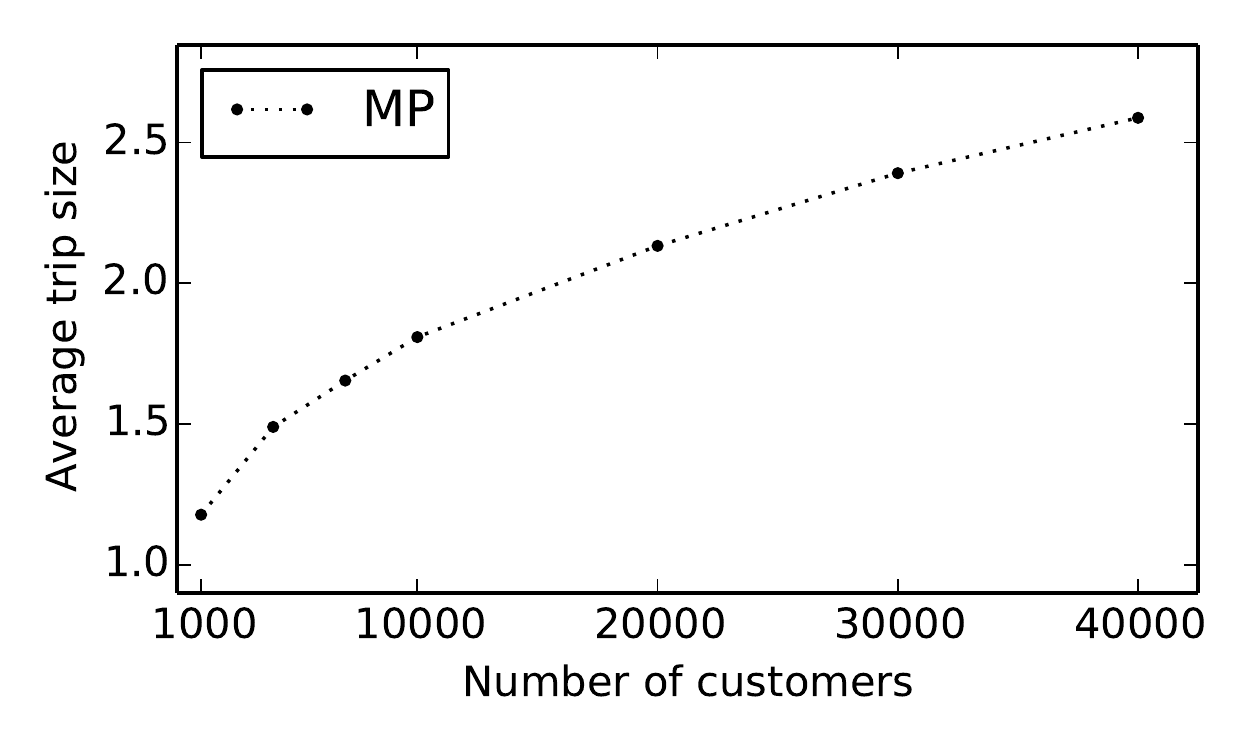} 
    \caption{Average customers per pick-up} 
    \label{fig:stats_tripsize} 
    \vspace{1ex}
  \end{subfigure}
  \begin{subfigure}[b]{0.5\linewidth}
    \centering
    \includegraphics[width=0.95\linewidth]{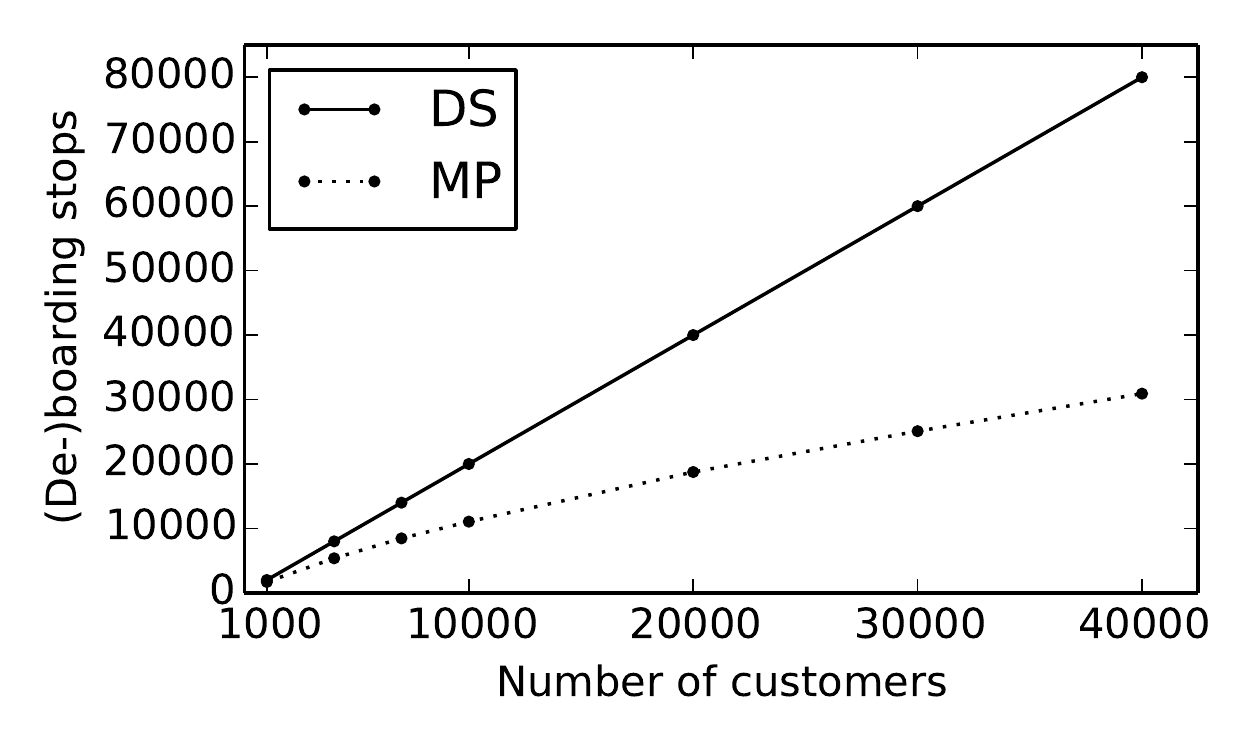} 
    \caption{Amount of boarding and alighting stops} 
    \label{fig:stats_stops} 
    \vspace{1ex}
  \end{subfigure}
  \caption{Comparison between meeting point and doorstep simulation concerning trip size. DS = Doorstep, MP = Meeting Points}
  \label{fig:comparison_tripsize} 
\end{figure}

Figure \ref{fig:groupsize} shows the group size histogram after the MP candidates selection step for four cases with different total number of customers (up to 40,000). While after the initial clustering phase nearly all groups have the maximum allowed size (in our case 11), the next step splits these groups into feasible subgroups. The result naturally correlates with the average number of customers per pick-up (Figure \ref{fig:stats_tripsize}). Generally, it can be stated that, with increasing number of customers, the portion of bigger groups increases as more people with similar itineraries and time schedules can be grouped together.
On the other hand, in our study area of Braunschweig, 4000 passengers are already sufficient for a majority of people to share their rides. This also inherently reduces the number of necessary boarding and de-boarding service stops for the vehicles (Figure \ref{fig:stats_stops}), since they have to stop only once for a group instead of stopping for every single customer. Naturally, the savings are higher when the demand is dense. With 5000 customers, the number of necessary stops is reduced by 33 \%, while it is halved at about 15000 customers.\\

After the vehicle routing optimization phase, several further statistics about actual vehicle usage and trip times can be derived (see tables \ref{tab:results_MP} and \ref{tab:results_DS} in the appendix). Figure \ref{fig:comparison_vehicles} gives an impression about the potential benefits for the operator when using meeting points (MP) instead of offering pick-ups at the doorstep (DS). While the effect is relatively small in low demand cases, the benefits are more significant with increasing number of customer requests. All plots in Figure \ref{fig:comparison_vehicles} show a similar trend. In the 10000 customers case, the savings in time, kilometre, and fleet size is up to 30\%. In addition, there are fewer dead kilometres (the distance travelled when the vehicle is without a passenger) in the MP case.

\begin{figure}[ht] 
  \begin{subfigure}[b]{0.5\linewidth}
    \centering
    \includegraphics[width=0.95\linewidth]{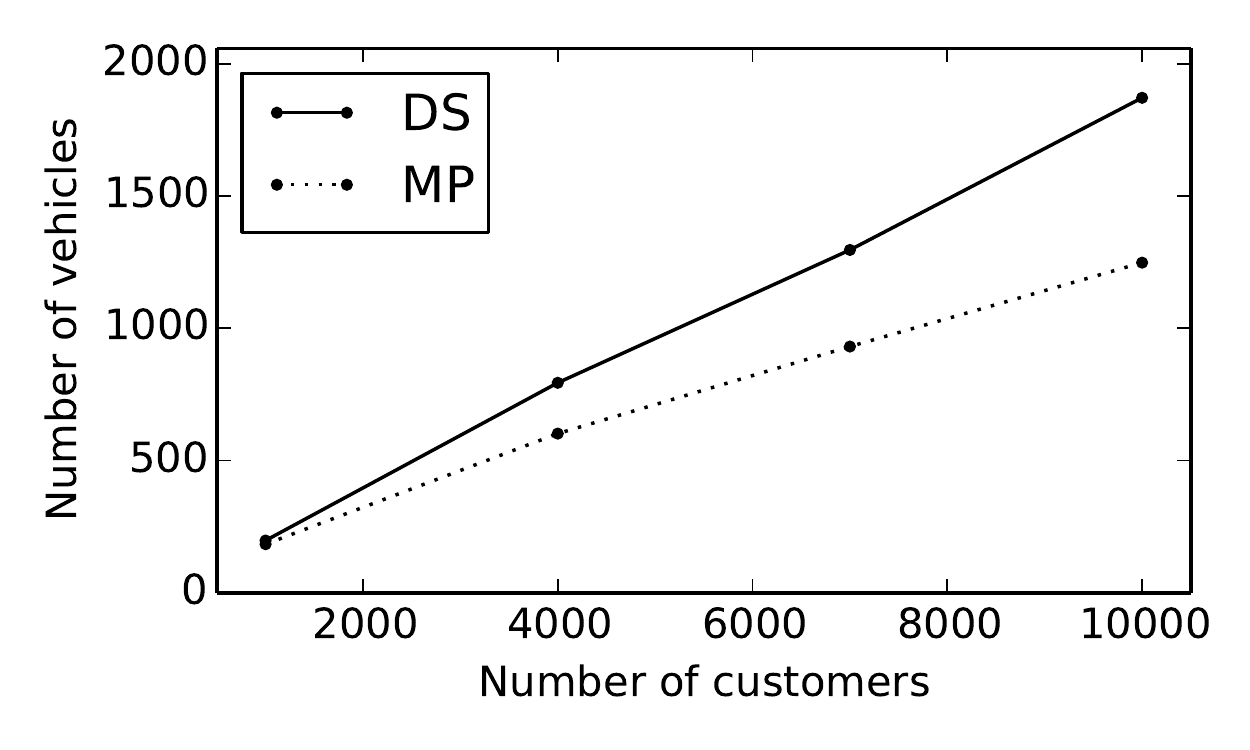} 
    \caption{Number of used vehicles} 
    \label{fig:stats_vehicles_no} 
    \vspace{1ex}
  \end{subfigure}
  \begin{subfigure}[b]{0.5\linewidth}
    \centering
    \includegraphics[width=0.95\linewidth]{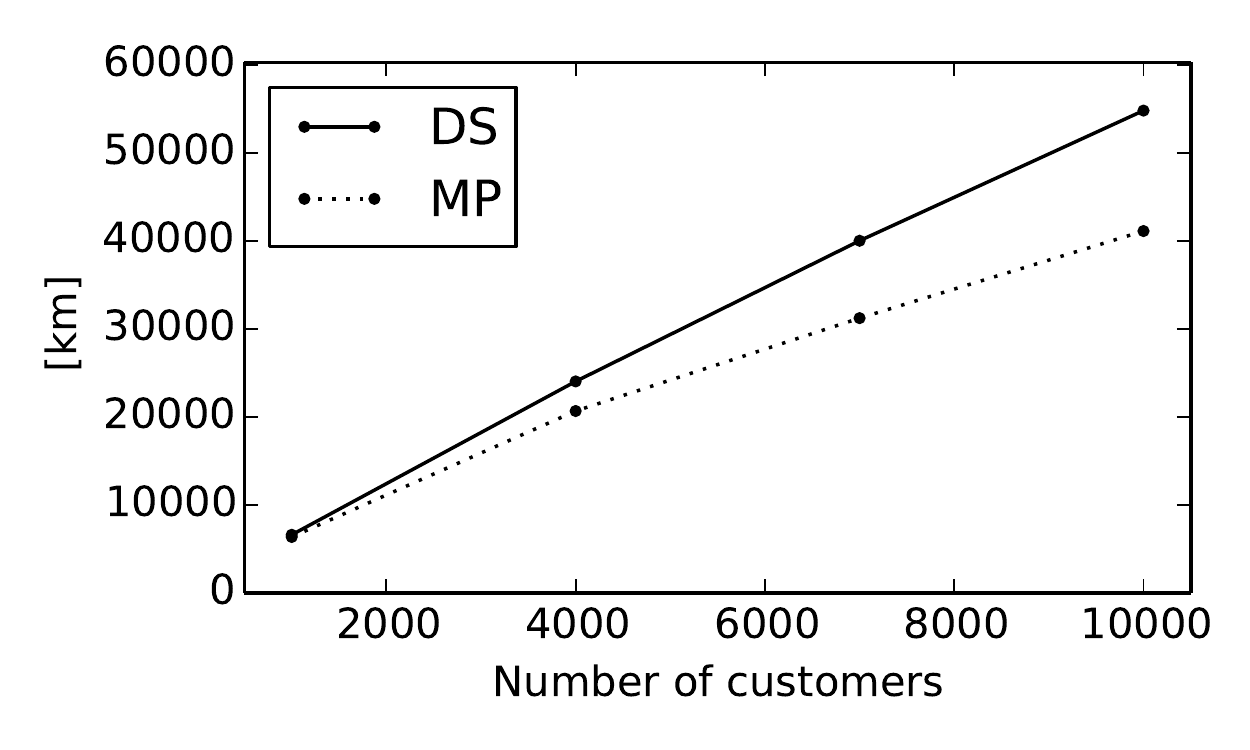} 
    \caption{Vehicle dead kilometers} 
    \label{fig:stats_vehicles_deadkm} 
    \vspace{1ex}
  \end{subfigure} 
  \begin{subfigure}[b]{0.5\linewidth}
    \centering
    \includegraphics[width=0.95\linewidth]{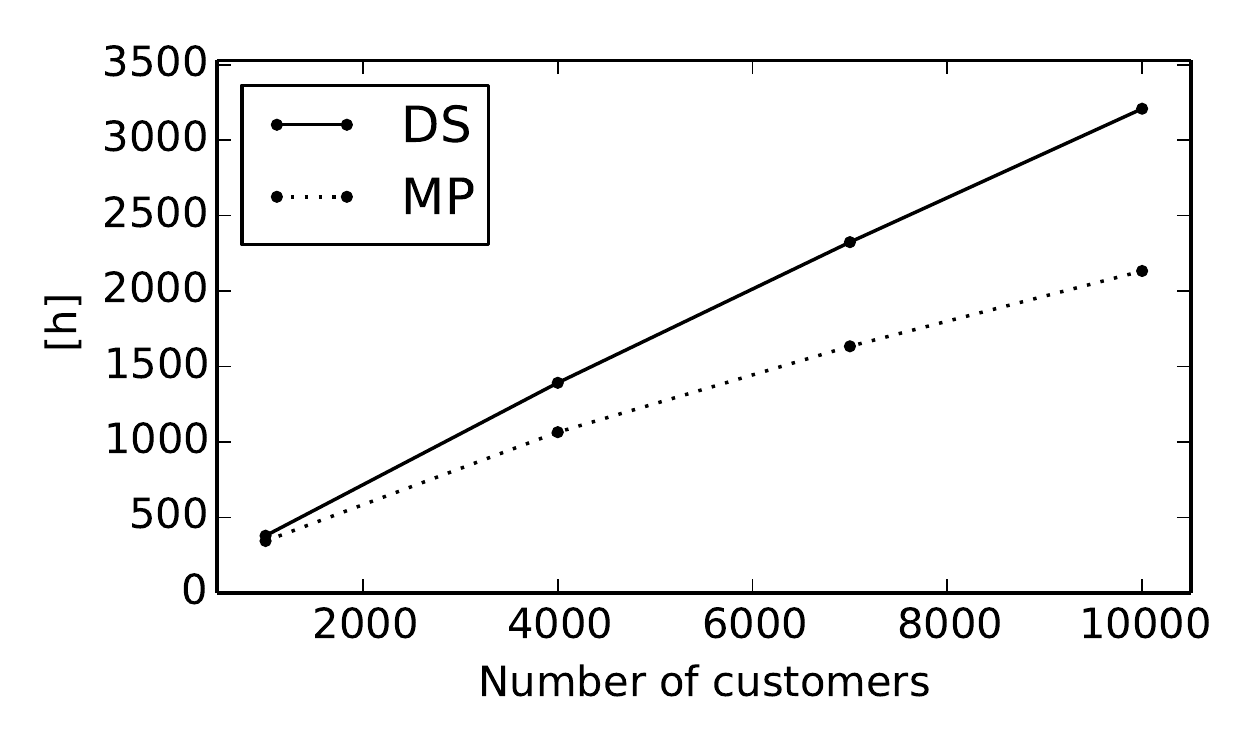} 
    \caption{Total vehicle hours} 
    \label{fig:stats_vehicles_h} 
    \vspace{1ex}
  \end{subfigure}
  \begin{subfigure}[b]{0.5\linewidth}
    \centering
    \includegraphics[width=0.95\linewidth]{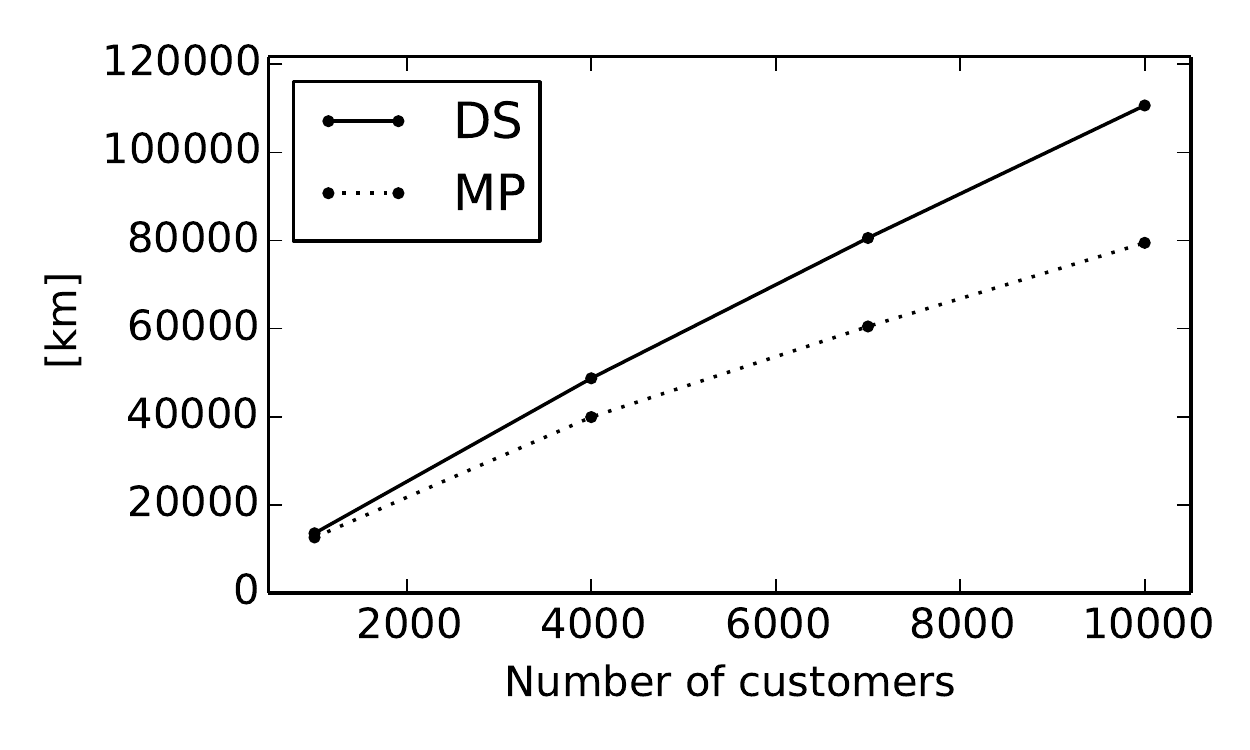} 
    \caption{Total vehicle mileage} 
    \label{fig:stats_vehicles_km} 
    \vspace{1ex}
  \end{subfigure}
  \caption{Comparison between meeting point and doorstep simulation concerning vehicle statistics (DS = Doorstep, MP = Meeting Points)}
  \label{fig:comparison_vehicles} 
\end{figure}

\begin{figure}[ht] 
  \begin{subfigure}[b]{0.5\linewidth}
    \centering
    \includegraphics[width=0.95\linewidth]{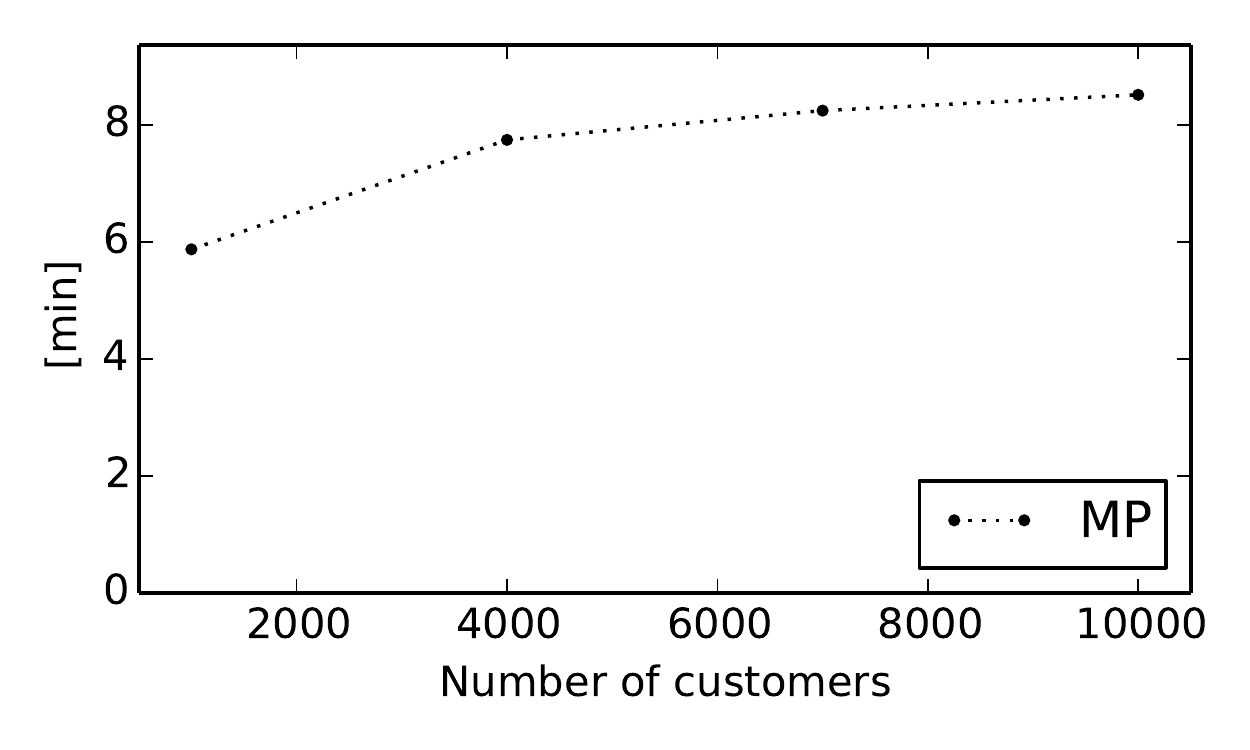} 
    \caption{Passenger average walking time} 
    \label{fig:stats_passenger_walk} 
    \vspace{1ex}
  \end{subfigure}
    \begin{subfigure}[b]{0.5\linewidth}
    \centering
    \includegraphics[width=0.95\linewidth]{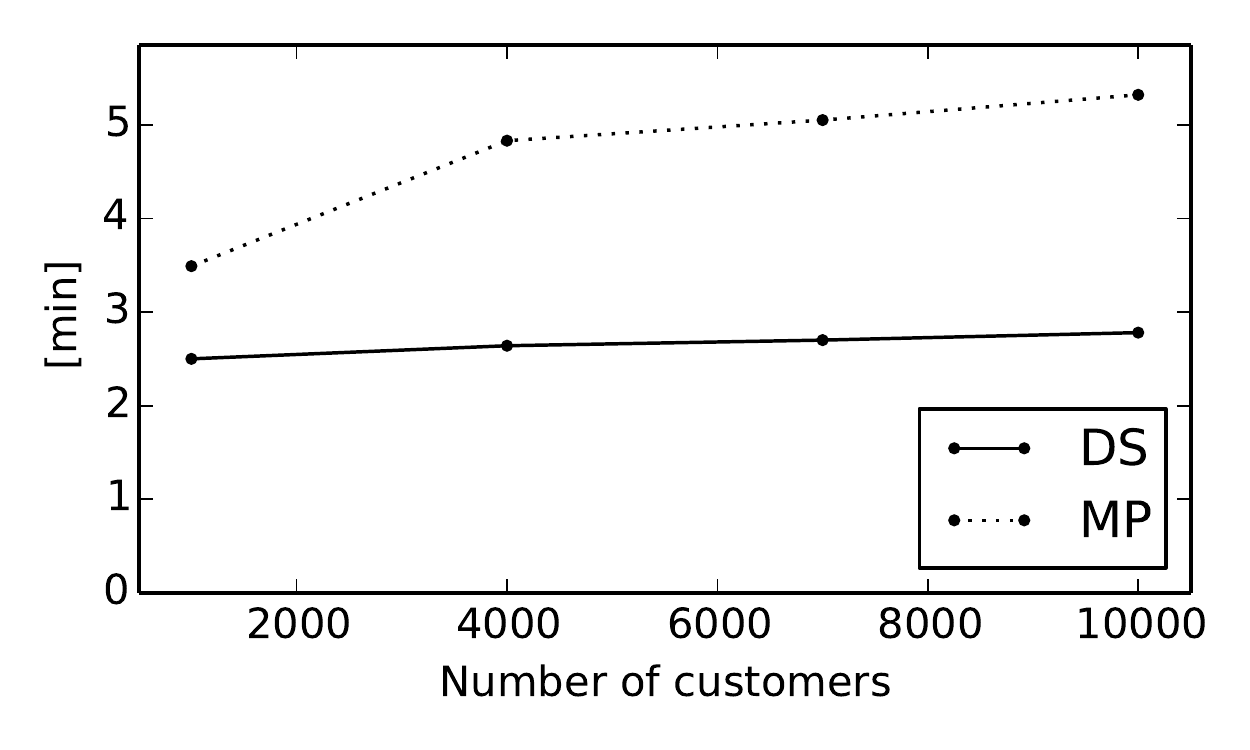} 
    \caption{Passenger average pickup wait time} 
    \label{fig:stats_passenger_wait} 
    \vspace{1ex}
  \end{subfigure}
  \begin{subfigure}[b]{0.5\linewidth}
    \centering
    \includegraphics[width=0.95\linewidth]{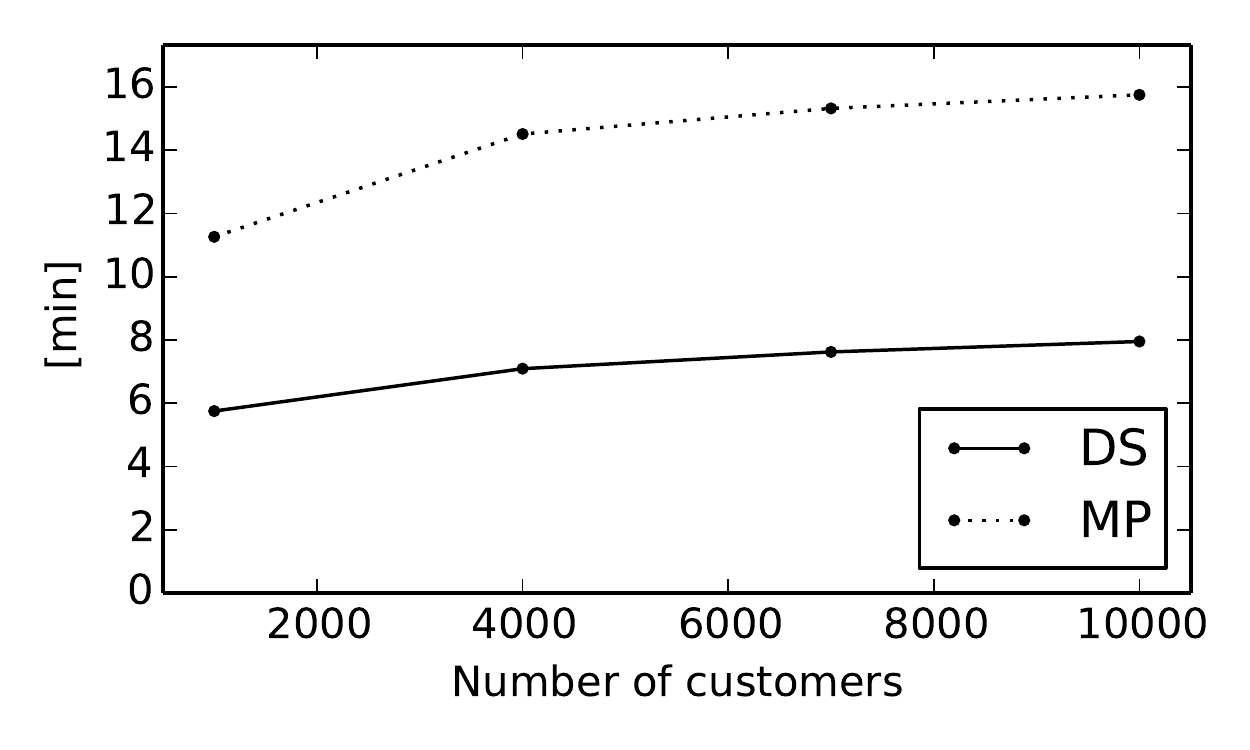} 
    \caption{Passenger average detour time} 
    \label{fig:stats_passenger_detour} 
    \vspace{1ex}
  \end{subfigure}

  \caption{Comparison between meeting point and doorstep simulation concerning passenger statistics (DS = Doorstep, MP = Meeting Points)}
  \label{fig:comparison_passengers} 
\end{figure}

\FloatBarrier
Figure \ref{fig:comparison_passengers}, on the other hand, shows the dimension of drawback for the passengers resulting from the usage of meeting points. Naturally, the walking time is an additional factor that has to be considered in the total travel times (Figure \ref{fig:stats_passenger_walk}). The average walking times in our example range from 6 to 8 minutes, including the walking time from the alighting point to the destination. For the doorstep case, the walking time is obviously zero. In addition to the walking time, the average waiting times at the meeting points are higher compared to the doorstep case (Figure \ref{fig:stats_passenger_wait}), since passengers likely have to wait for other fellow travellers. Here, the average waiting times for the pick-up are almost doubled when using meeting points, but the absolute values with about 3 to 5.5 minutes are comparably low. The total travel time differences between meeting and doorstep case can be seen in Figure \ref{fig:stats_passenger_detour}.

\section{Discussion} \label{sec:discussion}

In general, the results show benefits for operators of SDRT systems when using meeting and divergence points for the pick-up and drop-off procedure. Fewer vehicles, less mileage and a high reduction of necessary boarding and de-boarding stops can help reducing the operational costs, especially for cases where the demand is high.

However, these benefits for the operator come at the cost of some inconvenience for the customers as they have to spend additional time walking and waiting for fellow travellers. On the other hand, the reductions in operational costs may translate to a benefit for the customers if the travel price is subsequently reduced. This benefit can be expected to increase if more customers participate. In cases with a relatively high number of customers, the waiting and walking times do not significantly increase as the problem size increases (Figure \ref{fig:comparison_vehicles}). In contrast, the operators saving is consistently improving (Figure \ref{fig:comparison_passengers}).
This is due to the fact that more customers can be grouped with more flexibility, thereby reducing the distance to the closest meeting point within the cluster. The walking time in the 1000 customer case (Figure \ref{fig:stats_passenger_walk}) is only lower because of the many single traveller which are simply assigned to the closest available meeting point candidate. The additional walking and waiting times are in general lower than the difference in detour time (Figure \ref{fig:stats_passenger_detour}). Thus, the point-to-point transfer times in the MP case is faster than those in the DS case, which might be caused by the reduced total service time.

In the simulation of \cite{hall2008evaluation}, it was found that, in general, the DS case offers a better service for the customers, which can be confirmed by the results of this paper regarding travel time. However, they found no major differences in the results between the MP and the DS case and state that, according to their results, a door-to-door service can be offered without any noticeable loss in efficiency. This contradicts the findings of this paper, where an improved operational efficiency has been demonstrated (Figure \ref{fig:comparison_vehicles}). 

Despite performed on a ride-sharing system (which only allows one boarding and de-boarding stop per vehicle), the experiment by \cite{stiglic2015benefits} also concludes that the introduction of meeting points can improve a number of performance metrics, such as mileage savings and an increase in the number of matched riders. In their simulations, the total walking time is, on average, between 8 and 9 minutes, which is compareable to the finding of this paper. Furthermore, they state that the average trip time for matched riders increase by approximately 12\% due to the walking to or/and from a meeting point. In this paper, the travel time increase is up to 44\%, which can be attributed to the relatively low total travel times (17.5 minutes for the DS case, 25.3 minutes for the MP case on average). Since Braunschweig is a small city and congestion is not modelled, all nodes of the city network can be reached within a short time, and hence meeting and waiting times have a high impact on the result.

A drawback is the computation time of the presented workflow approach, especially for the vehicle routing optimization. While the instances are still solvable within a reasonable time for static demand (i.e.\ a planning problem), the computation time is still too large for real-time applications using dynamic requests.

\section{Conclusion}

Shared demand-responsive transportation systems offer the possibility to share travel costs among several passengers. For an efficient implementation, it is crucial to group people in a smart way to reduce detours and service stops. This paper introduces a workflow to include real-world meeting point locations in a SDRT solver, based on a partitioning of the demand. In a computational study, it has been demonstrated that the usage of meeting points in SDRT services can be beneficial for the operator in terms of vehicle usage, operation hours and mileage at the cost of increased travel times for the passengers, mainly due to walking and waiting times. In real-world applications, the recommended meeting points can offer some advantages concerning safe boarding and reliable localization. Further works include utilising real-world travel demand data and the design of an improved solver which can be used for real-time operation.

\setcounter{secnumdepth}{0}

\section{Acknowledgments}
This research has been supported by the German Research Foundation (DFG) through the Research Training Group SocialCars (GRK 1931), the Australian Research Council’s Linkage Projects funding scheme (project number LP120200130), the Universities Australia and the German Academic Exchange Service (DAAD) under the Australia-Germany Joint Research Co-operation Scheme.

\section{Appendix}

\subsection{Algorithms}
\label{sec:algorithms}

\noindent Note on notation and algorithms. Medium brackets are use to indicated
access to a certain element in the vector or a set e.g.~given a set $S$, $S[0]$
indicates an access to the first element. From this you can also notice that we
use zero-based indexing which is common in C like languages. In many algorithms
we used mathematical notation for what are common algorithmic procedures, this
allows algorithms to be more compact while still keeping same level of
information.

\begin{algorithm}
\caption{Clustering algorithm}
\label{alg:clustering}
\begin{algorithmic}[1]
    \Statex{Given a set of all requests $X$ and a maximum cluster size $k^{\text{max}}$.}
    \Statex{$\Rcal \gets \{ \}$}
    \Statex{$\Ccal \gets \{ \}$}
    \While{\Call{SizeOf}{$X$} $> 0$}\Comment{\Call{SizeOf}{$\cdot$} returns the number of elements in a set}
      \If{\Call{SizeOf}{$\Ccal$} $= 0$}
        \State{$\Ccal \gets \Ccal \cup \{\Call{PopFront}{$X$}\}$}\Comment{\Call{PopFront}{$X$} removes and returns first element}
      \EndIf{}
      \If{\Call{SizeOf}{$\Ccal$} $\gets k^{\text{max}}$}
        \State{$\Rcal \gets \Rcal \cup \{\Ccal\}$}
        \State{$\Ccal \gets \{ \}$}
      \EndIf{}
      \State{$x^* \gets \min\limits_{x \in X} d(\Ccal, x)$}\Comment{$\text{d}(\cdot, \cdot)$ is defined in Equation~\ref{eq:generalized_distance}}
      \State{$\Ccal \gets \Ccal \cup \{x^*\}$}
    \EndWhile{}
\end{algorithmic}
\end{algorithm}

\FloatBarrier
\newpage
Algorithm \ref{alg:meetingPoints} describes the procedure for the meeting point determination for a given demand cluster. The mentioned \textit{2-Combinations} function yields all possible paired combinations of a given set, e.g. \textit{2-Combinations(a,b,c,d) = [a-bcd, b-acd, c-abd, d-abc, ab-cd, ac-bd, ad-bc]}. The algorithm returns the optimal combination in terms of number of subgroups and summed squared walking distances.
\begin{algorithm}[htb!]
    \caption{Meeting Point Determination}\label{alg:meetingPoints}
    \begin{algorithmic}[1]
	\Procedure{FindCombination}{$c$} \Comment{Procedure is called for every cluster $c \in \Ccal$}
		\State{Given: Passenger group $c\gets\{\rho_{1},\rho_{2},\dots\}$}
		\State{$\Mcal \gets \Mcal_{\rho_{1}} \cap \Mcal_{\rho_{2}} \cap \dots$ \Comment{Find common meeting points}}
		\State{$\Dcal \gets \Dcal_{\rho_{1}} \cap \Dcal_{\rho_{2}} \cap \dots$ \Comment{Find common divergence points}}
		\State{$\Mcal_{TF}, \Dcal_{TF} \gets \{\}$}
		\If{$\mid \Mcal \mid \geq 1 \; \& \; \mid \Dcal \mid \geq 1$} \Comment{Check if at least one common MP and DP exists}
		    \For{$\mu \in \Mcal$} \Comment{Check temporal feasibility for common meeting points}
          \State{$t^{E}(\mu) \gets \max\limits_{\rho \in c}(t^{E}_{\rho}(\mu))$} \Comment{Earliest possible departure time at MP}
          \State{$t^{L}(\mu) \gets \min\limits_{\rho \in c}(t^{L}_{\rho}(\mu))$} \Comment{Latest possible departure time at MP}
		        \If{$t^{E}(\mu) \leq t^{L}(\mu)$} \Comment{Check time feasibility for the meeting point}
		            \State{$\gamma(\mu) \gets \sum_{\rho \in c} d^{W}(v^{+}_{\rho} \rightarrow \mu)^{2}$ \Comment{Calculate cost for this meeting point}}
		            \State{$\Mcal_{TF} \gets \Mcal_{TF} \cup \left(\gamma(\mu), \mu\right)$ \Comment{Time feasible - add to set}}
		        \EndIf{}
		    \EndFor
		    \State{Compute $\Dcal_{TF}$ likewise for divergence points}
		    \If{$\mid \Mcal \mid \geq 1 \; \& \; \mid \Dcal \mid \geq 1$} \Comment{Check if time feasible common MP and DP exist}
		        \State{$\gamma^{*}(\mu), \mu^{*} \gets \min(\Mcal_{TF})$ \Comment{Find $\mu$ with minimal cost}}
		        \State{$\gamma^{*}(\delta), \delta^{*} \gets \min(\Dcal_{TF})$ \Comment{Find $\delta$ with minimal cost}}
		        \State\Return $\Scal \gets (1, \gamma^{*}(\mu) + \gamma^{*}(\delta), \mu^{*}, \delta^{*}$) \Comment{Return combined cost, $\mu$ and $\delta$}
		    \EndIf{}
	    \Else \Comment{No common meeting and divergence point exists}
        \State{$\Scal_{Cur} \gets (|c|, \infty, \{\}, \{\})$} \Comment{Initialize current best solution}
	        \For{$c1, c2 \in$ 2-Combinations($c$)}\Comment{Iterate through possible pairwise combinations}
            \State{$\Scal_{1} \gets \Call{FindCombination}{$c1$}$}
            \State{$\Scal_{2} \gets \Call{FindCombination}{$c2$}$}
	            \State{$\Scal_{New} \gets \Scal_{1} \cup \Scal_{2}$}
	            \If{$\Scal_{New}$ better than $\Scal_{Cur}$} \Comment{Better = Less separate groups}
              \State{$\Scal_{Cur} \gets \Scal_{New}$}
	            \EndIf{}
          \EndFor{}
	    \EndIf{}
	    \State\Return $\Scal_{Cur}$
	\EndProcedure
	\end{algorithmic}
\end{algorithm}

\FloatBarrier

\begin{algorithm}[htb!]
    \caption{Recursive Alternative Meeting Point Search}\label{alg:meetingPointAlternatives}
    \begin{algorithmic}[1]
      \Statex{\textbf{Input}}
      \Statex{$\Mcal$}\Comment{Set of meeting points that all passengers of a trip can reach.}
      \Statex{$\Mcal_{C} \subset \Mcal$}\Comment{Set of meeting points determined by Algorithm~\ref{alg:meetingPoints}.}
      \Statex{$A_{R}$}\Comment{Travel time ratio matrix.}
      \Statex{$\sigma^{*}$}\Comment{Travel time ratio threshold }
      \Statex{\textbf{Return}}
      \Statex{$\Mcal_{C}$}\Comment{Set of meeting points to be considered }
      \Procedure{FindMPAlternatives}{$\Mcal, \Mcal_{C}$}
        \State{$\Rcal \gets \emptyset$\Comment{Initialize empty result set}}
          \For{$\alpha \in \Mcal \setminus \Mcal_{C}$}
              \State{$\Scal \gets \emptyset$\Comment{Initialize empty temporary set}}
              \For{$\beta \in \Mcal_{C}$}
                \State{$\Scal \gets \Scal \cup A[\alpha][\beta]$}
              \EndFor{}
              \State{$\Rcal \gets \Rcal \cup (\min(\Scal), \alpha)$\Comment{Get minimum value among all already considered points}}
          \EndFor{}
          \State{$\gamma \gets \max(\Rcal)$\Comment{Get maximum to choose point with highest minimal value}}
          \If{$\gamma[0] \geq \sigma^{*}$}\Comment{Check if value is above the threshold}
            \State{$\Mcal_{C} \gets \Mcal_{C} \cup \gamma[1]$\Comment{Add this point to the set of considered points}}
            \State{\Return\Call{FindMPAlternatives}{$\Mcal,\Mcal_{C}$}\Comment{Search for more points based on current setting}}
          \EndIf{}
        \State{\Return $\Mcal_{C}$}
      \EndProcedure{}
    \end{algorithmic}
\end{algorithm}

\FloatBarrier

\begin{algorithm}[htb!]
	\caption{Route neighbourhood search}\label{alg:route}
	\begin{algorithmic}[1]
    \Statex{\bf Input}
    \Statex{Initial route $\Rcal^k:=\{s^k_1,s^k_2,\dots,s^k_N\}$ where $s^k_i$ is the $i^{th}$ stop of vehicle $k$.}
    \Statex{$\Rcal^k_\text{best} \gets \Rcal^k$}
		\For{a pre-specified number of iterations}
		\State $\Rcal^{neighbour}_\text{best} \gets \Rcal^k$
		\For{$\gamma\in\{1,2,\dots,\text{neighbourhood size}\}$}
		\State Randomly choose a stop $s^k_i$ with $1<i<N$.
		\State Choose a random appropriate stop reinsertion spot
		\State $\Rcal^{neighbour}_{\gamma} \gets$ the new route after reinsertion and DP optimisation 
		\If{$J(\Rcal^{neighbour}_\gamma) < J(\Rcal^{neighbour}_\text{best})$} \Comment{$J(\cdot)$ objective function of a single vehicle.}
		\State $\Rcal^{neighbour}_\text{best} \gets \Rcal^{neighbour}_\gamma$
		\EndIf
		\EndFor
		\If{$J(\Rcal^{neighbour}_\text{best}) < J(\Rcal^k_\text{best})$}
		\State $\Rcal^k_\text{best} \gets \Rcal^{neighbour}_\text{best}$
		\EndIf
		\State $\Rcal^k \gets \Rcal^{neighbour}_\text{best}$ \label{alg:route_alwaysmove}
		\EndFor
	\end{algorithmic}
\end{algorithm}

\FloatBarrier

\begin{algorithm}[htb!]
	\caption{Trip allocation neighbourhood search}\label{alg:trip}
	\begin{algorithmic}[1]
    \Statex{\bf Input}
    \Statex $\Acal_k:=\{\tau_i\}$ \Comment{$\tau_i$ represents trip $i$ and $k$ is vehicle index}
		\Statex $\Acal^k_\text{best} \gets \Acal^k$, $\forall k$.
		\Statex $\Rcal^k_\text{best} \gets \Rcal^k$, $\forall k$.
		\For{a pre-specified number of iterations}
		\State $\Acal^{neighbour,k}_\text{best} \gets \Acal^k$, $\forall k$.
		\State $\Rcal^{neighbour,k}_\text{best} \gets \Rcal^k$, $\forall k$.
		\For{$\gamma\in\{1,2,\dots,\text{neighbourhood size}\}$}
		\State Randomly choose a trip $\tau_i\in\cup_k \Acal^k$. 
    \State $k \gets \eta_{1}$ \Comment{$\eta_1$ is vehicle index of the chosen trip.}
    \State $\beta \gets \{k \mid \Acal^k \neq \emptyset \} \setminus \{\eta_1\} \cup \min\{k \mid \Acal^k \neq \emptyset \}$.
    \State Reinsert the trip to a random vehicle $k\in\beta$. \Comment{greedy insertion}
    \State $k \gets \eta_{2}$ \Comment{$\eta_2$ is the vehicle index being inserted.}
		\State $\Acal^{neighbour,k}_{\gamma} \gets$ the new trip allocation after reinsertion, $\forall k$
		\State $\Rcal^{neighbour,k}_{\gamma} \gets \Rcal^k$, $\forall k\notin\{\eta_1,\eta_2\}$
		\State Re-optimise $\Rcal^{\eta_1}$ and $\Rcal^{\eta_2}$ using Algorithm \ref{alg:route}.
		\State $\Rcal^{neighbour,\eta_1}_{\gamma} \gets$ $\Rcal^{\eta_1}$
		\State $\Rcal^{neighbour,\eta_2}_{\gamma} \gets$ $\Rcal^{\eta_2}$
		\If{$\sum_k J(\Rcal^{neighbour,k}_\gamma) < \sum_k J(\Rcal^{neighbour,k}_\text{best})$}
		\State $\Acal^{neighbour,k}_\text{best} \gets \Acal^{neighbour,k}_\gamma$, $\forall k$.
		\State $\Rcal^{neighbour,k}_\text{best} \gets \Rcal^{neighbour,k}_\gamma$, $\forall k$.
		\EndIf
		\EndFor
		\If{$\sum_k J(\Rcal^{neighbour,k}_\text{best}) < \sum_k J(\Rcal^k_\text{best})$}
		\State $\Acal^k_\text{best} \gets \Acal^{neighbour,k}_\text{best}$, $\forall k$.
		\State $\Rcal^k_\text{best} \gets \Rcal^{neighbour,k}_\text{best}$, $\forall k$.
		\EndIf
		\State $\Acal^k \gets \Acal^{neighbour,k}_\text{best}$, $\forall k$.
		\State $\Rcal^k \gets \Rcal^{neighbour,k}_\text{best}$, $\forall k$.
		\EndFor
	\end{algorithmic}
\end{algorithm}

\FloatBarrier
\newpage
\subsection{Tabular results}

\begin{table}[htb]
	\centering
	\caption{The results for the MP case} \label{tab:results_MP}
		\begin{tabular}{|c|c|c|c|c|}
			\hline
			No.\ of riders & 1000 & 4000 & 7000 & 10000 \\
			\hline\hline
			Final objective function value	& 498,835 & 2,319,368 & 4,086,774 & 3,815,542 \\
			Total vehicle kms* 		& 12,594 & 39,915 & 60,485 & 79,470 \\
			Total vehicle hours 	& 344 & 1065 & 1634 & 2133 \\
			No.\ of vehicles used* 	& 184 & 602 & 931 & 1248 \\
			Total dead kms 			& 6,357 & 20,668 & 31,222 & 41,118 \\
			Total idle hours 		& 36 & 91 & 134 & 161 \\
			Total empty-idle hours 	& 33 & 84 & 121 & 145 \\
			\hline
			Trips average pick-up waiting time (minutes) 	& 2.05 & 1.73 & 1.54 & 1.39 \\
			Trips average detour time (minutes) 			& 2.17 & 2.42 & 2.56 & 2.4 \\
			Passengers average pick-up waiting time (minutes) 	& 3.49 & 4.83 & 5.05 & 5.32 \\
			Passengers average detour time (minutes) 			& 11.26 & 14.51 & 15.32 & 15.75 \\
			Passengers average walk time (minutes)				& 5.87 & 7.74 & 8.24 & 8.51 \\
			\hline
			Route optimisation computation time (hours) & 0.94 & 3.12 & 4.51 & 5.30 \\
			\hline
		\end{tabular}%
\end{table}

\FloatBarrier

\begin{table}[htb]
	\centering
	\caption{The results for the DS case} \label{tab:results_DS}
		\begin{tabular}{|c|c|c|c|c|}
			\hline
			No.\ of riders & 1000 & 4000 & 7000 & 10000 \\
			\hline\hline
			Final objective function value	& 413,294 & 1,632,910 & 2,701,750 & 3,895,434 \\
			Total vehicle kms* 		& 13,536 & 48,723 & 80,570 & 110,655 \\
			Total vehicle hours 	& 379 & 1392 & 2324 & 3208 \\
			No.\ of vehicles used* 	& 198 & 794 & 1296 & 1871 \\
			Total dead kms 			& 6,603 & 24,032 & 40,016 & 54,806 \\
			Total idle hours 		& 32 & 91 & 136 & 160 \\
			Total empty-idle hours 	& 28 & 71 & 106 & 119 \\
			\hline
			Passengers average pick-up waiting time (minutes) 	& 2.50 & 2.64 & 2.70 & 2.77 \\
			Passengers average detour time (minutes) 			& 5.75 & 7.09 & 7.62 & 7.95 \\
			\hline
			Route optimisation computation time (hours) & 3.15 & 9.79 & 16.50 & 25.80 \\
			\hline
		\end{tabular}%
\end{table}

\bibliographystyle{informs2014trsc}
\bibliography{mybibfile}

\end{document}